\documentclass[draft]{agujournal2019}
\usepackage{url} 
\usepackage[inline]{trackchanges} 
\usepackage{soul}
\usepackage{gensymb}
\usepackage{amsmath}

\draftfalse

\journalname{JGR: Space Physics}

\begin{document}

\title{Constraining Europa's subsolar atmosphere with a joint analysis of HST spectral images and Galileo magnetic field data}

\authors{S. Cervantes \affil{1}, J. Saur \affil{1}}

\affiliation{1}{Institute of Geophysics and Meteorology, University of Cologne, Germany}

\correspondingauthor{S. Cervantes}{jcervant@uni-koeln.de}



\begin{keypoints}
\item We combine HST spectral images and Galileo magnetometer data to constrain the density and location of water vapor in Europa's atmosphere

\item We simulate the plasma interaction for the Galileo E12 flyby with a three-component atmosphere: global $\mathrm{O_2}$, stable confined $\mathrm{H_2O}$, and a plume 

\item Using 50\% of $\mathrm{O_2}$ and 50\% to 75\% of $\mathrm{H_2O}$ column densities from \citeA{roth2021stable} yields magnetic field signatures consistent with both observations
\end{keypoints}

%
%

%
%


\begin{abstract}
We constrain Europa's tenuous atmosphere on the subsolar hemisphere by combining two sets of observations: oxygen emissions at 1304 \AA~and 1356 \AA~from Hubble Space Telescope (HST) spectral images, and Galileo magnetic field measurements from its closest encounter, the E12 flyby. We describe Europa's atmosphere with three neutral gas species: global molecular ($\mathrm{O_2}$) and atomic oxygen (O), and localized
water ($\mathrm{H_2O}$) present as a near-equatorial plume and as a stable distribution concentrated around the subsolar point on the moon's trailing hemisphere. Our combined modelling based on the ratio of OI 1356 \AA~to OI 1304 \AA~emissions from \citeA{roth2021stable} and on magnetic field data allows us to derive constraints on the density and location of $\mathrm{O_2}$ and $\mathrm{H_2O}$ in Europa's atmosphere. We demonstrate that $50\%$ of the $\mathrm{O_2}$ and between $50\%$ and $75\%$ of the $\mathrm{H_2O}$ abundances from \citeA{roth2021stable} are required to jointly explain the HST and Galileo measurements. These values are conditioned on a column density of $\mathrm{O}$ close to the upper limit of $6 \times10^{16}~\mathrm{m}^{-2}$ derived by \citeA{roth2021stable}, and on a strongly confined stable $\mathrm{H_2O}$ atmosphere around the subsolar point. Our analysis yields column densities of $1.2 \times10^{18}~\mathrm{m}^{-2}$ for $\mathrm{O_2}$, and $1.5 \times10^{19}~\mathrm{m}^{-2}$ to $2.2 \times10^{19}~\mathrm{m}^{-2}$  at the subsolar point for $\mathrm{H_2O}$. Both column densities however still lie within the uncertainties of \citeA{roth2021stable}. Our results provide additional evidence for the existence of a stable $\mathrm{H_2O}$ atmosphere at Europa.
\end{abstract}


%
%

%


%
%
%
%

\section{Introduction}

Europa is thought to harbor a global liquid water  ($\mathrm{H_2O}$) ocean under its icy surface \cite{carr1998evidence, khurana1998induced, kivelson2000galileo}, and is therefore a prominent candidate in the search for extraterrestrial life. Previous observations of water vapor in the form of transient plumes rising above Europa's surface \cite{roth2014transient} might carry the possibility to probe the ocean water that is ejected into the atmosphere, and the upcoming ESA's JUICE \cite{grasset2013jupiter} and NASA's Europa Clipper missions \cite{howell2020nasa} have initiated further interest to better understand this moon's atmosphere, its interior, and its plasma environment. 

Molecular oxygen ($\mathrm{O_2}$) was the first constituent to be detected in Europa's atmosphere \cite{hall1995detection}, but a stable $\mathrm{H_2O}$ component, in contrast to the sporadic plumes, remained undetected for a long time. \citeA{roth2021stable} analyzed a set of Hubble Space Telescope (HST) spectral images, and provided the first evidence of a persistent $\mathrm{H_2O}$ distribution in the central sunlit trailing hemisphere of the moon. This same region was traversed by the Galileo spacecraft in 1997 along its E12 flyby, and the magnetometer on board measured the magnetic field as the spacecraft approached the moon on its closest encounter. 

The primary means of detecting Europa's neutral gas environment is via emission of its atomic constituents. \citeA{hall1995detection} performed the first observations of the moon's atmosphere using HST observations, and the ultraviolet (UV) spectrum revealed emissions at 1304 \AA~and 1356 \AA. The ratio of atomic oxygen emission at these two wavelengths, $r_\gamma = \mathrm{OI}~1356$ \AA $/ \mathrm{OI}~1304$ \AA, yielded a value of 1.9, which implied electron impact dissociative excitation of $\mathrm{O_2}$ as the emission process. Later studies \cite <e.g.>[]{hall1998far, roth2016europa} presented additional sets of HST UV images of Europa's atmosphere, and their measured ratios $r_\gamma$ were consistently larger than 1. These results supported the conclusion that Europa's atmosphere is dominated by $\mathrm{O_2}$. Years later, \citeA{roth2014transient} reported surpluses of hydrogen Lyman-$\alpha$ and $\mathrm{OI}~1304$ \AA~emissions near Europa's south pole from HST images. Their results were interpreted as a local atmospheric inhomogeneity, consistent with an active water plume as a source. The lack of detection of these emissions in other observations suggested varying plume activity of intermittent nature.

Recently, \citeA{roth2021stable} inspected the radial profile of the oxygen emission ratio $r_\gamma$ for several HST observations at different orbital locations of Europa. A major finding was that for the trailing side visits, $r_ \gamma$ systematically decreased from the limb towards the disk center. This profile was shown to be in agreement with  an $\mathrm{H_2O}$-dominated atmosphere concentrated around the subsolar point, and an $\mathrm{O_2}$-dominated atmosphere elsewhere. Furthermore, the reduced oxygen emission ratio on the disk center was found to be consistent within uncertainties among the four trailing side visits, obtained between 1999 and 2015. However, the source of this persistent $\mathrm{H_2O}$ atmosphere could not be unambiguously identified, as the values calculated by \citeA{roth2021stable} are approximately two orders of magnitude larger than the predicted $\mathrm{H_2O}$ column densities for sputtering and sublimation of water ice at Europa's surface temperature \cite{shematovich2005surface, smyth2006europa, plainaki2013exospheric, vorburger2018europa}.

Several models have been developed to describe the moon's atmosphere and to better constrain its generation process. \citeA{shematovich2005surface} and \citeA{smyth2006europa} used a Monte Carlo (MC) technique for the water group species to determine the atmospheric compositional structure and gas escape rates. \citeA{plainaki2010neutral} and \citeA{plainaki2012role} performed an MC calculation for the generation of Europa's atmosphere and incorporated sputtering information from laboratory measurements.  \citeA{teolis2017plume} also implemented an MC model, and assumed a water plume source with multiple organic and nitrile species, in addition to sputtering, radiolysis, and other surface processes. \citeA{vorburger2018europa} modelled the formation of Europa's atmosphere via an MC code and considered sputtering by ions and electrons, as well as sublimation for some species.

Europa orbits Jupiter at the outer edge of the inner magnetosphere and is constantly overtaken by the corotating Jovian plasma. Close to the moon, ionization and collisions within Europa's atmosphere modify the plasma flow around it and generate magnetic field perturbations. Over eight close flybys between 1996 and 2000, the instruments on board Galileo measured local field and plasma perturbations, and hence provided a tool to probe Europa's neutral gas environment. Various numerical simulations following different approaches have been performed in order to match the spacecraft observations and to understand the plasma interaction with Europa and its atmosphere. Such models range from two-fluid codes \cite<e.g.>[]{saur1998interaction}, to single-fluid magnetohydrodynamic (MHD) \cite <e.g.>[]{kabin1999europa, schilling2007time, schilling2008influence, blocker2016europa} and multi-fluid MHD models \cite <e.g.>[]{rubin2015self, harris2021multi}, and hybrid codes \cite <e.g.>[]{arnold2019magnetic, arnold2020plasma}. These numerical simulations have been employed to estimate plasma production and neutral loss rates, constrain the atmosphere distribution, explore the properties of a subsurface ocean, and study the effect of localized water plumes.

In this work, we present a parametrization of Europa's subsolar atmosphere and provide constraints on the column densities and location of the neutral $\mathrm{O_2}$ and $\mathrm{H_2O}$ by combining two datasets: (a) the observed profile of the oxygen emission ratio from HST spectral images by \citeA{roth2021stable}, and (b) magnetic field measurements collected by the Galileo magnetometer (MAG) for its E12 flyby. First, we vary the abundances of $\mathrm{O_2}$, O, and $\mathrm{H_2O}$ calculated by \citeA{roth2021stable} to derive several possible distributions that fit the emission ratio profile, all within the uncertainties of the observations. Next, we use these distributions in a three-dimensional MHD code and simulate Europa's interaction with the Jovian plasma. These results allow us to identify the densities that are the most consistent both with the HST and MAG data. Finally, we consider uncertainties in certain parameters of the atmospheric and MHD model and assess the robustness of our results. 
  
This paper is organized as follows: in Section 2, we present the neutral atmosphere model and compute the emission intensities, and in Section 3 we describe the single-fluid MHD model for the plasma interaction. In Section 4, we present our derived oxygen emission ratio profiles for several assumed neutral gas distributions, and in Section 5 we show the respective MHD simulations. In Section 6, we perform a parameter study for different 
$\mathrm{H_2O}$ and electron properties, and we also discuss our findings with respect to the plasma environment and Europa's neutral atmosphere. Finally, Section 7 summarizes the most important results.

\section{Atmosphere Model and Emission Rates}\label{sec:atmosphere model and emission rates}

We assume a model of Europa's neutral atmosphere consisting of three species: $\mathrm{O_2}$, $\mathrm{O}$, and $\mathrm{H_2O}$, and we simulate the respective electron-excited oxygen emissions. The goal is to reproduce the observed radial profile of the oxygen emission ratio from \citeA{roth2021stable} using a simplified description with as few parameters as possible.

\subsection{Atmosphere Model}\label{subsec:atmosphere model}

For the three neutral gas species, we consider exponentially decreasing radial distributions with the column densities estimated by \citeA{roth2021stable}. The $\mathrm{O_2}$ distribution is considered global as this molecule does not stick to the surface, as $\mathrm{H_2O}$ does, or thermally escape Europa's gravity, as $\mathrm{H_2}$ does \cite{johnson2009composition, mcgrath2009observations, plainaki2018towards}. Therefore, it is the dominant species in Europa's atmosphere \cite{hall1995detection}, and it accumulates approximately uniformly over the moon \cite {mcgrath2009observations, bagenal2020space}. Previous modelling studies \cite <e.g.>[] {saur1998interaction, schilling2007time,jia2018evidence,arnold2019magnetic} have considered an upstream-downstream asymmetry in the $\mathrm{O_2}$ atmosphere. However, in this work we deliberately omit this asymmetry and keep the $\mathrm{O_2}$ distribution as simple as possible in order to better demonstrate the effects of the localized $\mathrm{H_2O}$ on the plasma interaction. The scale height of the global $\mathrm{O_2}$ is fixed to 150 km, as considered in previous modelling studies \cite <e.g.>[]{saur1998interaction, schilling2007time}, and similar to the best fit OI 1356 \AA~scale height from \citeA{roth2016europa}. With an $\mathrm{O_2}$ column density of $N_\mathrm{O_2} = 2.47 \times 10^{18}~\mathrm{m}^{-2}$, we get a surface number density of $n_{\mathrm{O_2},0} = 1.64 \times 10^{13}~\mathrm{m}^{-3}$. The number density of the radially symmetric $\mathrm{O_2}$ is given by:

\begin{equation}
n_\mathrm{O_2}(h) = n_\mathrm{O_2,0}~\mathrm{exp}\left(-\frac{h}{H_\mathrm{O_2}}\right),
\label{eq:o2 density}
\end{equation}

\noindent with scale height $H_\mathrm{O_2}$ and altitude $h = r - R_E$ above the surface, with Europa's radius $R_E = 1569~\mathrm{km}$.

As a second constituent, we consider atomic $\mathrm{O}$ produced through the dissociation of the molecular oxygen. Similar to $\mathrm{O_2}$, the abundance of $\mathrm{O}$ is also described by an exponential decrease. In line with \citeA{roth2016europa}, we assume a 2 times larger scale height for the lighter atomic $\mathrm{O}$, i.e., $H_\mathrm{O} =$ 300 km. With the derived upper limit for the $\mathrm{O}$ abundance from \citeA{roth2021stable}, equal to $6 \times 10^{16}~\mathrm{m}^{-2}$, the surface number density of atomic $\mathrm{O}$ is $n_{\mathrm{O},0} = 2 \times 10^{11}~\mathrm{m}^{-3}$. It must be emphasized that while atomic $\mathrm{O}$ is included in our atmospheric model to reproduce the observed profile of the oxygen emission ratio, it is not taken into account in the MHD modelling (Section \ref{sec:mhd}), as the maximum $\mathrm{O}/\mathrm{O_2}$ mixing ratio of 0.03 \cite{roth2021stable} makes it too dilute to impact the plasma interaction. 

In accordance with the results of \citeA{roth2021stable}, we assume an $\mathrm{H_2O}$ distribution strongly concentrated around the subsolar point in the trailing hemisphere, described by the following equation:

\begin{equation}
n_\mathrm{H_2O}(h, \alpha) = n_\mathrm{H_2O,0}~\mathrm{cos}^\beta(\alpha)~\mathrm{exp}\left(-\frac{h}{H_\mathrm{H_2O}}\right),
\label{eq:h2o density}
\end{equation}

\noindent where $\alpha$ is the angle to the subsolar point. $\mathrm{H_2O}$ freezes on contact with the icy surface, limiting its abundance in the atmosphere. Hence, the exponent $\beta$ is introduced in equation (\ref{eq:h2o density}) to restrict the spatial distribution. The resulting $\mathrm{H_2O}$ atmosphere is highly localized with a maximum at the subsolar point and is frozen on the nightside of Europa. 

The maximum dayside temperature at Europa's surface is 
132 K \cite{spencer1999temperatures}, and therefore we assume an $\mathrm{H_2O}$ scale height of 46 km. The column density is $N_\mathrm{H_2O} = 2.95 \times 10^{19}~\mathrm{m}^{-2}$ \cite{roth2021stable} at the subsolar point, which results in a surface number density of $n_{\mathrm{H_2O},0} = 6.41 \times 10^{14}~\mathrm{m}^{-3}$. The subsolar point is located at a longitude of $217.5\degree$ W (between the anti-Jovian meridian and the trailing hemisphere apex) and a latitude of $1\degree$ N as extracted from the Solar System SPICE kernel. 

In addition, \citeA{jia2018evidence} has provided in-situ evidence of a water plume on Europa from the magnetic field and plasma wave observations for the Galileo E12 flyby. Therefore, we also include the effect of a plume on the plasma interaction, and incorporate in our atmospheric model an analytical form for the density profile of the plume. We use the following description similar to \citeA{blocker2016europa}:

\begin{equation}
n_{\mathrm{pl}}(r, \tilde{\theta}) = n_\mathrm{pl,0} \cdot \mathrm{exp} \left[-\left(\frac{r-R_E}{H_\mathrm{pl}}\right) - \left(\frac{\tilde{\theta}}{H_\theta} \right)^2\right],
\label{eq:o2_model_plume}
\end{equation}

\noindent where $n_\mathrm{pl,0}$ is the surface number density of the neutral gas in the center of the plume, $H_\mathrm{pl}$ is the scale height, $\tilde{\theta}$ is the angular distance from the center of the plume, and $H_\theta$ is the opening angle. The numerical values used are: $n_\mathrm{pl,0} = 3\times10^{15}~\mathrm{m}^{-3}$, $H_\mathrm{pl} = 150$ km, and $H_\theta = 3^{o}$. The angular distance $\tilde{\theta}(\theta,\phi)$ from the vector pointing from the center of Europa to the center of the plume at the surface is given by:

\begin{equation}
\tilde{\theta}(\theta,\phi) = \mathrm{arccos} \left[ \mathrm{sin}(\theta)\mathrm{sin}(\theta_\mathrm{ap})\mathrm{cos}(\phi - \phi_\mathrm{ap}) + \mathrm{cos}(\theta)\mathrm{cos}(\theta_\mathrm{ap})  \right].
\label{eq:o2_model_plume_part2}
\end{equation}

\noindent with the spherical coordinates of the plume center $\theta_\mathrm{ap}$ and $\phi_\mathrm{ap}$. Similar to \citeA{jia2018evidence}, the base of the plume is located at $245\degree$ W and $5\degree$ S. In addition, the plume is tilted with respect to the radial direction by $15\degree$ towards the east and $25\degree$ towards the south. 

The coordinate system employed to describe the water plume in our atmospheric model requires a more detailed discussion. We consider the center of Europa as the origin for the definition of the angular distance $\tilde{\theta}$, analogous to the approach of \citeA{saur2008evidence}, \citeA{roth2014transient}, \citeA{blocker2016europa}, and \citeA{blocker2018mhd}. In contrast, other studies use the footpoint of the plume at the surface as the origin \cite <e.g.>[]{jia2018evidence, arnold2019magnetic}. Physically, the Europa-centered approach describes a wider plume for similar $H_\theta$ outgassing over a wider area, and is consistent with localized heat spots or porous near surface structures. In contrast, the plume-centered approach depicts a narrower plume being ejected from a single point at the surface, and provides a representation of features such as short cracks. However, the plume-centered description generates a singularity at the origin of the coordinate system (i.e. the source of the plume). The Europa-centered approach does not pose this singularity at the surface of the moon and can be fully resolved numerically.

The two above-mentioned descriptions generate two distinct plume models and are therefore not equivalent. This implies that, despite employing the same numerical value for the opening angle, the water plume modelled in \citeA{roth2014transient} and \citeA{blocker2016europa} with $H_\theta = 15\degree$ does not possess the same width as the one considered in \citeA{jia2018evidence} and \citeA{arnold2019magnetic} with $\theta_p = 15\degree$ (c.f. equation 4 of \citeA{jia2018evidence}), since the opening angles are defined differently in both cases. 

\begin{figure}[]
\centering
\noindent\includegraphics[width=0.65\textwidth]{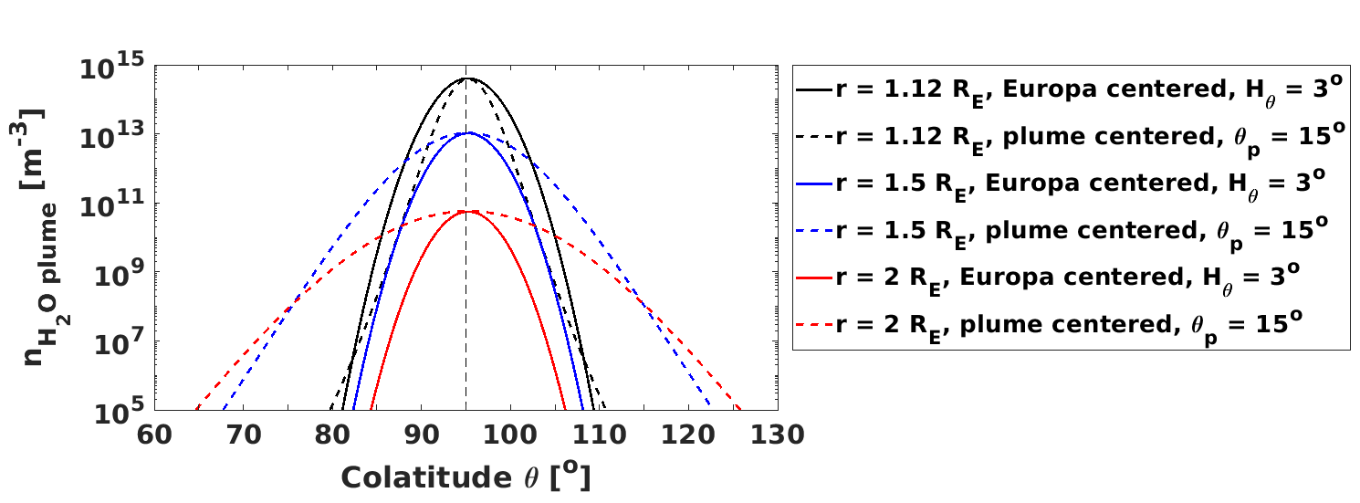}
\caption{Water plume neutral density for three different radial distances as a function of colatitude $\theta$. The solid lines correspond to the Europa-centered approach, while the dashed lines indicate the plume-centered description. $H_\theta$ represents the opening angle in the Europa-centered approach, whereas $\theta_p$ represents the opening angle in the plume-centered description (c.f. equation 4 of \citeA{jia2018evidence}). The vertical black dashed line indicates the colatitude of the footpoint of the plume at $\theta_{\mathrm{ap}} = 95\degree$.}
\label{fig:neutral-densities-plume}
\end{figure}

In this work, we take the center of Europa as the origin for our description of the angular distance. We adjust $H_\theta$ to  $3\degree$ in order to match the plume width at the altitude of closest approach of the E12 flyby (196 km, equivalent to $r=1.12~R_E$) with the one from \citeA{jia2018evidence}, who employed an opening angle $\theta_p = 15\degree$. Figure~\ref{fig:neutral-densities-plume} presents the $\mathrm{H_2O}$ plume neutral density calculated with both descriptions for three different radial distances as a function of colatitude. For $r=1.12~R_E$, the two approaches effectively yield a comparable width ($< 2\degree$ of difference).

\subsection{Emission Rates}\label{subsec:emission rates}

We compute the emission rates produced by electron impact excitation of Europa's neutral atmosphere at two specific wavelengths: 1304 \AA~and 1356 \AA. We assume a thermal electron population of 20 eV \cite{sittler1987io} plus a 250 eV suprathermal population \cite{johnson2009composition} with a $5\%$ mixing ratio \cite{bagenal2015plasma}. In accordance with \citeA{roth2021stable}, we consider an electron density of 160 $\mathrm{cm}^{-3}$. The collisional excitation rates $f_{n,\lambda}(T_e)$ at a wavelength $\lambda$ are given as an integral over the Maxwell-Boltzmann distribution $f_{\mathrm{Max}}$, the electron velocity $v(E)$, and the energy-dependent cross sections $\sigma_{n,\lambda}(E)$ for the collisions between the exciting electrons and the neutral species $n$ according to:  
 
\begin{equation}
f_{n,\lambda}(T_e)[\mathrm{m}^{3}~\mathrm{s}^{-1}] = \int_{E_t}^\infty f_{\mathrm{Max}}(E, T_e)~ \sigma_{n,\lambda}(E)~v(E)~\mathrm{d}E,
\label{eq:emission-rate}
\end{equation}

\noindent where $E_t$ is the energy of the excitation threshold. For our computation of the emission rates, we set $E_t$ to 14 eV as in \citeA{hartkorn2017structure}. The electron impact excitation cross sections are based on the laboratory measurements of OI 1304 \AA~and OI 1356 \AA~emission intensities by \citeA{doering1989absolute}, \citeA{kanik2001electron}, \citeA{kanik2003electron}, and \citeA{makarov2004kinetic}.  The local volume emission rates $i_{n,\lambda}$ are, in turn, calculated by multiplying the density of the neutral atmospheric gas with the density of the impinging electrons and the excitation rates, as follows:

\begin{equation}
i_{n,\lambda}[\mathrm{m}^{-3}~\mathrm{s}^{-1}] = n_e~n_n~f_{n,\lambda}(T_e).
\label{eq:local-emission-rate}
\end{equation}

\noindent The intensity $I_\lambda$ in Rayleigh at a specific wavelength $\lambda$ is then computed by integrating the local intensities over the line of sight:

\begin{equation}
I_{\lambda}[R] = 10^{-10}\sum_n\int_\mathrm{los} i_{n,\lambda}~\mathrm{d}s.
\label{eq:brightness}
\end{equation} 

We additionally calculate average intensities $I_{\lambda,\mathrm{av}}$ across the images in $0.025~R_E$ concentric rings around the disk center for both wavelengths, as follows:

\begin{equation}
I_{\lambda,\mathrm{av}}[R] = \frac{\int_0^{2\pi}\int_{r_1}^{r_2}I_\lambda~r~\mathrm{d}r\mathrm{d}\theta}{\pi(r^2_2-r^2_1)},
\label{eq:brightness-average}
\end{equation}

\noindent where $r_1$ and $r_2$ are the radii of the inner and outer circles limiting a concentric ring, respectively. Finally, the radial profile of the oxygen emission ratio $r_\gamma$ is obtained by dividing the averaged OI 1356 \AA~intensity by the averaged OI 1304 \AA~intensity in all pixels within the respective concentric rings, similar to \citeA{roth2021stable}.

\section{MHD Plasma Model}\label{sec:mhd}

In order to describe the plasma interaction with Europa's atmosphere, we apply a three-dimensional single-fluid MHD model, based on that of \citeA{duling2014consistent} and also employed by \citeA{blocker2016europa} and \citeA{blocker2018mhd} to describe Europa's and Io's plasma interaction, respectively. Our simulations self-consistently calculate the magnetic field and bulk plasma properties. With the model results, we constrain the $\mathrm{H_2O}$ atmosphere by comparing observed and modelled magnetic field perturbations near Europa.

\subsection{Geometry and Model Equations}
\label{subsec:model equations}

We use a Cartesian and a spherical coordinate system, both with their origin in the center of the moon. The Cartesian system is the EPhiO system where the $x$ axis points along the direction of the corotational plasma flow, the $y$ axis corresponds to the Jupiter-Europa vector, and the $z$ axis is parallel to Jupiter's spin axis. The spherical coordinate system is characterized by the radius $r$, the colatitude $\theta$ measured from the positive $z$ axis, and the longitude $\phi$ measured from the positive $y$ axis towards the negative $x$ axis.

Our single-fluid MHD model consists of one evolution equation for each of the following four plasma variables: magnetic field $\mathbf{B}$, plasma bulk velocity $\mathbf{v}$, plasma mass density $\rho$, and internal energy density $e$. The equations read:

\begin{equation}
    \frac{\partial \rho}{\partial t} + \nabla \cdot \left( \rho \mathbf{v} \right) = \left( P - L \right) m_i,
\label{eq:continuity}
\end{equation}

\begin{equation}
    \frac{\partial \mathbf{v}}{\partial t} + \left(\mathbf{v} \cdot \nabla \right) \mathbf{v} = -\frac{1}{\rho}\nabla p + \frac{1}{\rho\mu_0} \left( \nabla \times \mathbf{B} \right) \times \mathbf{B} - \left(\frac{Pm_i}{\rho} + \nu_n \right) \mathbf{v}, 
\label{eq:velocity}
\end{equation}

\begin{equation}
    \frac{\partial \mathbf{B}}{\partial t} = \nabla \times \left( \mathbf{v} \times \mathbf{B} \right),
\label{eq:induction}
\end{equation}

\begin{equation}
    \frac{\partial}{\partial t} \left( \frac{e}{\rho} \right) + \left( \mathbf{v} \cdot \nabla \right) \frac{e}{\rho} = -\frac{p}{\rho} \nabla \cdot \mathbf{v} + \frac{1}{2} v^2 \left( \frac{Pm_i}{\rho} + \nu_n \right) - e \left(\frac{Lm_i}{\rho^2} + \frac{\nu_n}{\rho} \right),
\label{eq:energy}
\end{equation}
    
\noindent with ion mass $m_i$, plasma production and loss rates $P$ and $L$, respectively, vacuum permeability $\mu_0$, ion-neutral collision frequency $\nu_n$, and plasma thermal pressure $p$, which is related to the internal energy density through $e = \frac{3}{2}p$. The plasma production and loss rates and the collision frequency specify various physical processes and their quantitative expressions are provided in the next section.

For the upstream magnetospheric plasma we use an average ion mass $\tilde{m}_i = 18.5$ u and an effective ion charge $z_c$ = 1.5 \cite{kivelson2004magnetospheric}, as in previous studies of the Europa-plasma interaction \cite <e.g.>[]{blocker2016europa,arnold2019magnetic,arnold2020plasma}. The upstream plasma mass density can be written as: $\rho_0 = \tilde{m}_in_e/z_c$ with the electron number density $n_e$. Finally, the upstream internal energy density is given by: $e_0 =\frac{3}{2}n_0k_B(T_e + T_i)$ with the background ion density $n_0 = \rho_0/\tilde{m}_i$.

Our upstream magnetospheric parameters are similar to those of \citeA{jia2018evidence}, who also modelled the plasma interaction around Europa for the Galileo flyby E12. We consider a bulk velocity of 100 $\mathrm{km~s^{-1}}$ in the corotation direction. The Jovian background magnetic field is determined by excluding the perturbed values of the Galileo magnetometer data around 10 min of closest approach, performing a linear fit, and then extracting the fitted magnetic field values at closest approach, which results in $\mathbf{B}_0 = (78,0,-395)$ nT. Based on Galileo's Plasma Wave Spectrometer (PWS) measurements, the upstream electron number density is set to 500 cm$^{-3}$ \cite{kurth2001plasma}, as derived from the upper hybrid resonance emissions. The ion and electron temperatures read $k_BT_{i} = k_BT_{e} = 100$ eV \cite{kivelson2004magnetospheric}, resulting in an upstream plasma mass density and internal energy density of $6.166 \times 10^9$ u m$^{-3}$ and $16.02 \times 10^{-9}~\mathrm{J~m^{-3}}$, respectively.

\subsection{Plasma Sources and Losses}\label{subsec:sources and losses}

According to \citeA{saur1998interaction}, the dominant ionization process in Europa's atmosphere is electron impact ionization, which is more than one order of magnitude larger than photoionization. Therefore, in our model the neutral atmosphere and plume are only ionized by electron impacts, and two ionospheric singly charged ion populations with masses $m_\mathrm{O_2^+} = 32$ u and $m_\mathrm{H_2O^+} = 18$ u are produced. The ion production rates for O$_2$ and H$_2$O are calculated by multiplying the respective neutral density by a given ionization rate, in analogy to \citeA{blocker2016europa}, \citeA{jia2018evidence}, and \citeA{arnold2019magnetic}. We adopt constant electron impact ionization rates of $f_{\mathrm{imp}} = 2 \times 10^{-6}~\mathrm{s}^{-1}$ for both $\mathrm{O_2}^+$ and $\mathrm{H_2O}^+$  production, within the range derived by \citeA{smyth2006europa}, and analogous to the values employed by \citeA{arnold2019magnetic} and \citeA{arnold2020plasma}.  

Dissociative recombination between ions and electrons is the main loss process in our model. We account for the loss of ionospheric O$_2^+$ and H$_2\mathrm{O}^+$ with the recombination rate coefficients $\alpha_\mathrm{rec}$ (in  $\mathrm{m}^3~\mathrm{s}^{-1}$) given by \citeA{schunk2009ionospheres}:

\begin{equation}
\alpha_\mathrm{rec, O_2^+}(T_e) = 2.4 \times 10^{-13}\left(\frac{300}{T_e}\right)^{0.7},
\label{eq:recombination_o2}
\end{equation}

\begin{equation}
\alpha_\mathrm{rec, H_2O^+}(T_e) = 1.03 \times 10^{-9}~T_e^{-1.111}.
\label{eq:recombination_h2o}
\end{equation}

\noindent For the calculation of $\alpha_{rec}$, we use an ionospheric electron temperature $T_e$ of 0.5 eV. In analogy to the approach of \citeA{duling2014consistent}, \citeA{blocker2016europa}, and \citeA{blocker2018mhd}, we avoid that the plasma number density $n = \rho/m_i$ decreases below the background ion density $n_0$ by adopting the expression for the loss rate:

\begin{equation}
L = \begin{cases}
\alpha_{\mathrm{rec}}n\left(n-n_0 \right) &\text{for}~n>n_0\\
0 &\text{for}~n<n_0
\end{cases}
\label{eq:recombination}
\end{equation}

\noindent from \citeA{saur2003ion}. 

The exchange of momentum between the plasma and Europa's atmosphere is modelled through the ion-neutral collision frequency:

\begin{equation}
\nu_n = \sigma_n v_0 n_n,
\label{eq:collisions}
\end{equation}

\noindent similar to \citeA{duling2014consistent}. Equation (\ref{eq:collisions}) is a function of the ion-neutral collision cross section $\sigma_n$, typical plasma bulk velocity $v_0$, and the number density $n_n$ of $\mathrm{O_2}$ and $\mathrm{H_2O}$ molecules in the atmosphere. We employ an O$_2$ cross section of $2 \times 10^{-19}~\mathrm{m}^{-2}$ as in \citeA{saur1998interaction} and an H$_2$O cross section of $8 \times 10^{-19}~\mathrm{m}^{-2}$ following equations (A2) to (A7) from \citeA{kriegel2014ion}. Two different mechanisms are included in the total momentum transfer cross sections: induced dipole ion-molecule interactions
and charge exchange.

\subsection{Electromagnetic Induction in a Subsurface Water Ocean}\label{subsec:electromagnetic induction}

Due to the $\sim$10$\degree$ tilt of Jupiter's magnetic moment with respect to its spin axis, the $x$ and $y$ components of the Jovian background magnetic field vary periodically at Europa's location. This results in an inducing field with the 11.1 h synodic rotation period of Jupiter. The time-varying inducing background magnetic field, in units of nT, is given analytically as a function of system III longitude by \cite{schilling2007time}:

\begin{equation}
B_{0,x}(\lambda_\mathrm{III}) = -84~\mathrm{sin(\lambda_{III}-200\degree)},
\label{eq:b_ind_x}
\end{equation}

\begin{equation}
B_{0,y}(\lambda_\mathrm{III}) = -210~\mathrm{sin(\lambda_{III}-200\degree)}.
\label{eq:b_ind_y}
\end{equation}

\noindent In comparison to the strong variations of the other two field components, $B_{0,z}$ is about an order of magnitude smaller \cite{seufert2011multi}. This time-varying inducing background field drives currents in Europa's conductive subsurface water ocean, and therefore generates a time-varying induced dipolar magnetic field \cite{khurana1998induced, saur2010induced}. Considering a spatially homogeneous inducing magnetic field and a radially symmetric ocean, the induced field is dependent on the thickness, the conductivity, and the depth of the ocean beneath the surface. In accordance with \citeA{schilling2007time} and \citeA{blocker2016europa}, we assume an ocean that is 100 km thick and lies 25 km below the surface, with an electric conductivity of $\sigma = 0.5~\mathrm{S}~\mathrm{m}^{-1}$. The time-variable induced field within the subsurface ocean is included in the inner boundary conditions at the surface of Europa as discussed in section \ref{subsec:boundary conditions}.

\subsection{Numerical Solution Process}\label{subsec:numerics}
In order to solve the differential equations (\ref{eq:continuity}) to (\ref{eq:energy}), we utilize a modified version of the three-dimensional publicly available ZEUS-MP MHD code. This is a multi-physics, massively parallel, message-passing open source code first developed by \citeA{stone1992zeus1} and \citeA{stone1992zeus2}, which solves the single-fluid, ideal MHD equations in three dimensions. ZEUS-MP uses a finite-difference staggered-mesh approach and applies a second-order accurate, monotonic advection scheme. The solution is computed by the code time forward and the time step is controlled by the Courant-Friedrichs-Lewy criterion. In addition, ZEUS-MP combines the Constrained Transport algorithm with the Method of Characteristics (MOC-CT) for the treatment of Alfv\'en waves. ZEUS-MP algorithms are described in detail in \citeA{stone1992zeus2}, \citeA{stone1992zeus1}, and \citeA{hayes2006simulating}.

We employ a spherical grid with $160 \times 360 \times 360~(r, \theta, \phi)$ cells. The angular resolution is equidistant in $\theta$ and $\phi$ with $\Delta \theta = 0.5\degree$ and $\Delta \phi = 1\degree$. The radial resolution is not equidistant and we increase the radial grid spacing by a factor of 1.026 from cell to cell, between the inner ($r = 1~R_E$) and the outer boundary ($r = 20~R_E$), which results in a resolution at the surface equal to 13 km.

\subsection{Boundary Conditions}\label{subsec:boundary conditions}
The two boundary areas of our simulation domain are the inner sphere at $r = 1~R_E$ and the outer sphere at $r = 20~R_E$. At the outer boundary we apply open boundary conditions for the four MHD variables $\rho$, $\mathbf{v}$, $\mathbf{B}$, and $e$. At the upstream region ($\phi \leq 180\degree$) the inflow method is used, while at the downstream region ($\phi > 180\degree$) the outflow method is applied. At the inner boundary, i.e., Europa's icy surface, plasma particles are assumed to be absorbed. Therefore we utilize open boundary conditions for $\mathbf{v}$, $\rho$, and $e$ by an outflow method. The radial component of the plasma bulk velocity is set to $v_r \leq 0$ everywhere on the surface, as no plasma flows out of it. Furthermore, Europa's insulating icy surface also inhibits any electric currents penetrating it. \citeA{duling2014consistent} derived boundary conditions for the magnetic field ensuring there is no radial electric current. In addition, the boundary condition also includes any time-dependent internal potential fields from below the surface, e.g. due to induction in an ocean beneath the nonconducting icy crust.



\section{Oxygen Emission Ratio Profile}
\label{sec:oxygen emission ratio}

We now quantitatively investigate Europa's neutral gas environment and present our atmospheric distributions along with their two-dimensional emission patterns. We also show the respective simulated oxygen emission ratios and compare them with the observed radial profile derived by \citeA{roth2021stable} from HST spectral images.

Our description of Europa's atmosphere consists of four free parameters: column densities of the three species under consideration (i.e. $N_\mathrm{O_2}$, $N_\mathrm{H_2O}$, and $N_\mathrm{O}$), and degree of confinement of the stable $\mathrm{H_2O}$ as described by the exponent $\beta$ of the cosine term in equation~\ref{eq:h2o density}. Our aim is to derive constraints on the abundances of $\mathrm{O_2}$ and $\mathrm{H_2O}$, and on the spatial extent of the stable $\mathrm{H_2O}$, taking into account the uncertainty on the $\mathrm{O}$ column density. The density of atomic oxygen is observationally less constrained than molecular oxygen, and typically, only upper limits for $N_\mathrm{O}$ have been derived \cite <e.g.>[]{hansen2005cassini,shematovich2005surface,roth2021stable}. Therefore, we assume the interval ranging from the absence of atomic oxygen ($N_\mathrm{O} = 0$) to the upper bound $N_\mathrm{O} = 6\times10^{16}~\mathrm{m}^{-2}$ calculated by \citeA{roth2021stable}. Concerning the degree of confinement of the stable $\mathrm{H_2O}$, we assume values of $\beta$ extending from 2 to 10.

\subsection{Emission Ratio Without Atomic Oxygen}
\label{subsec:oxygen emission ratio-no oxygen}

\begin{figure}[]
\centering
\noindent\includegraphics[width=0.5\textwidth]{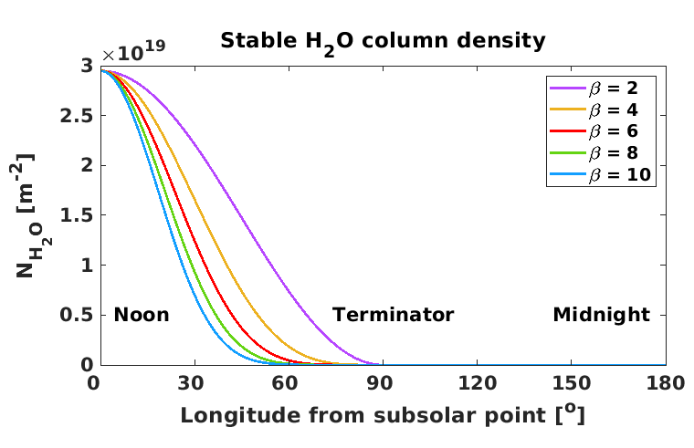}
\caption{Degree of confinement of the $\mathrm{H_2O}$ component. Column density distribution as a function of longitude from the subsolar point for different values of the exponent $\beta$.}
\label{fig:h2o-column-density-beta}
\end{figure}

We start by considering the lowermost limit of the $N_\mathrm{O}$ uncertainty range ($N_\mathrm{O} = 0$), and calculate the emission ratio for an $\mathrm{O_2}$ together with a stable $\mathrm{H_2O}$ atmosphere, with $N_\mathrm{O_2}$ and $N_\mathrm{H_2O}$ as given in \citeA{roth2021stable}, and with the abundance of $\mathrm{H_2O}$ relative to $\mathrm{O_2}$ equal to 12. Regarding the stable $\mathrm{H_2O}$, we examine the three following values of $\beta$: 2, 6, and 10, characterizing a weakly, moderately, and strongly confined $\mathrm{H_2O}$ distribution around the subsolar point, respectively. Figure~\ref{fig:h2o-column-density-beta} shows the $\mathrm{H_2O}$ column density as a function of longitude from the subsolar point for $\beta$ ranging from 2 to 10. All the distributions peak at the subsolar point, at 12 Local Time (LT), but decrease at a different rate away from it. For example, in the least confined $\mathrm{H_2O}$ atmosphere ($\beta = 2$), the column density reaches half of its maximum value ($N_\mathrm{H_2O} = 1.5 \times 10^{19}~\mathrm{m}^{-2}$) at $45\degree$ away from the subsolar point, whereas in the most localized case ($\beta = 10$), such an $\mathrm{H_2O}$ column density is observed $22\degree$ away from it. In addition, the rate of decrease of the stable $\mathrm{H_2O}$ abundance differs less markedly between the cases with the largest exponents.
 
\begin{figure}[b!]
\centering
\noindent\includegraphics[width=1\textwidth]{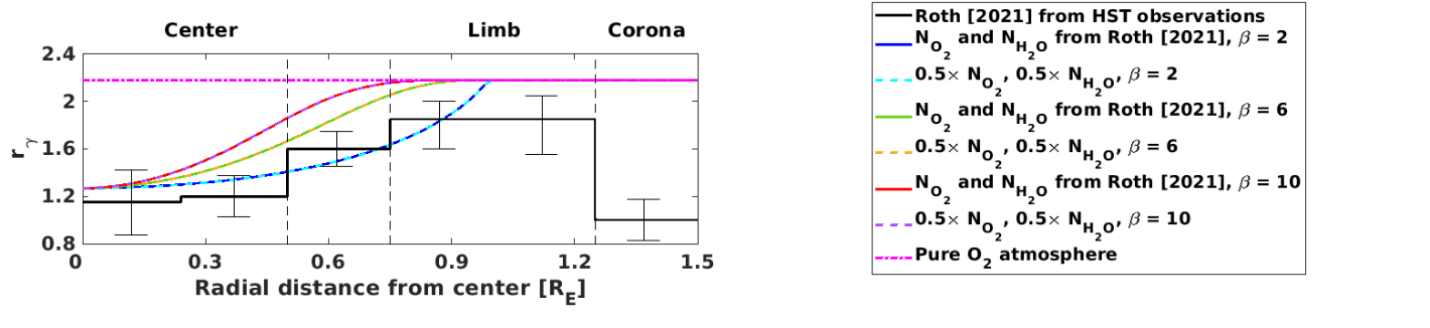}
\caption{Radial profile of the oxygen emission ratio of OI 1356 \AA~to OI 1304 \AA~for HST observations (black histogram) and for our simulated combined $\mathrm{O_2}$ and $\mathrm{H_2O}$ atmosphere models with varying degree of confinement given by the exponent $\beta$. The $\mathrm{O_2}$ and $\mathrm{H_2O}$ column densities of the profiles in dashed and solid lines correspond to 50\% and 100\% of the values from \citeA{roth2021stable}, respectively.}
\label{fig:oxygen-emission-ratio-without-O}
\end{figure}

The radial profiles of the observed and simulated oxygen emission ratios $r_\gamma$ are depicted as solid lines in Figure~\ref{fig:oxygen-emission-ratio-without-O}. All the profiles exhibit a minimum at the disk center, where the stable $\mathrm{H_2O}$ is confined, and a gradual increase in $r_\gamma$ towards the limb, due to the contribution of the global $\mathrm{O_2}$ to the total emission. Beyond $1~R_E$, our modelled profiles stay constant at $r_\gamma \sim 2.18$ and are consistent with a pure $\mathrm{O_2}$ atmosphere, whereas the observed $r_\gamma$ decreases due to the limb emission by $\mathrm{O}$. Only the radial profile of the model with the least confined $\mathrm{H_2O}$ atmosphere ($\beta = 2$) fits the error bars, while the profiles for $\beta = 6$ and $\beta = 10$ do not match the observed ratio within its uncertainties, except in the innermost bin.

We then calculate the emission ratio for the same atmospheric distributions, i.e. $\mathrm{O_2}$ combined with a stable $\mathrm{H_2O}$, except that the column densities of both neutrals are reduced by 50\% with respect to the values from \citeA{roth2021stable}, keeping the mixing ratio $N_\mathrm{H_2O}/N_\mathrm{O_2} = 12$. The scale heights of both species do not vary, i.e. $H_\mathrm{O_2}$ and $H_\mathrm{H_2O}$ are kept constant. Rather, the surface number density is recalculated, as the column density is given by the product of the assumed scale height and the surface number density. The corresponding oxygen emission ratio profiles are shown as dashed lines in Figure~\ref{fig:oxygen-emission-ratio-without-O}. Reducing the column density of both $\mathrm{O_2}$ and $\mathrm{H_2O}$ by 50\% yields the same $r_\gamma$ as with the original abundances. Neglecting the $10^{-10}$ factor for the conversion in units Rayleigh and the averaging in concentric rings from equations \ref{eq:brightness} and \ref{eq:brightness-average}, this can be shown as follows:

\begin{equation}
r_{\gamma,50\%} = \frac{\mathrm{OI~1356~\AA}}{\mathrm{OI~1304~\AA}} = \frac{\int_{\mathrm{los}}n_e \times 0.5n_{\mathrm{O_2}} \times f_{\mathrm{O_2},\mathrm{1356~\AA}}~\mathrm{d}s + \int_{\mathrm{los}} n_e \times 0.5n_{\mathrm{H_2O}} \times f_{\mathrm{H_2O},\mathrm{1356~\AA}}~\mathrm{d}s}{\int_{\mathrm{los}}n_e \times 0.5n_{\mathrm{O_2}} \times f_{\mathrm{O_2},\mathrm{1304~\AA}}~\mathrm{d}s + \int_{\mathrm{los}} n_e \times 0.5n_{\mathrm{H_2O}} \times f_{\mathrm{H_2O},\mathrm{1304~\AA}}~\mathrm{d}s} = r_{\gamma,100\%},
\label{eq:rgamma-ratio}
\end{equation}
 
\noindent where the subindex in $r_\gamma$ indicates the percentage of the column densities from \citeA{roth2021stable}. The 0.5 factors multiplying the $\mathrm{O_2}$ and $\mathrm{H_2O}$ neutral densities in the numerator and the denominator in equation~\ref{eq:rgamma-ratio} cancel out, and hence, our simulated $r_\gamma$ with reduced column densities is identical to the $r_\gamma$ with the original values from \citeA{roth2021stable}. This statement holds for all different abundances of $\mathrm{O_2}$ and $\mathrm{H_2O}$, as long as both of them are decreased by the same percentage, and thus, the mixing ratio $N_\mathrm{H_2O}/N_\mathrm{O_2}$ is held constant. Therefore, in the absence of atomic oxygen, $r_\gamma$ only provides constraints on the spatial extent of the stable $\mathrm{H_2O}$, but not on the column densities of $\mathrm{O_2}$ and $\mathrm{H_2O}$ in the atmosphere.
 
\subsection{Emission Ratio With Atomic Oxygen}
\label{subsec:oxygen emission ratio-with oxygen}

\begin{figure}[]
\includegraphics[width=0.75\textwidth]{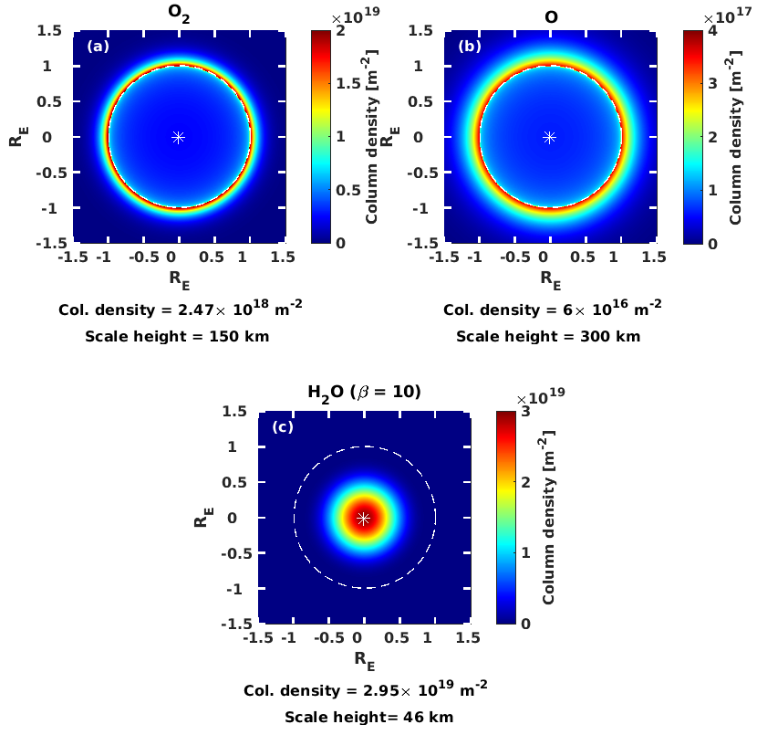}
\caption{Line-of-sight integrated column density maps of the individual $\mathrm{O_2}$, $\mathrm{O}$, and stable $\mathrm{H_2O}$ (with $\beta = 10$) distributions in the trailing hemisphere. The subsolar point is located at the center of the disk and is indicated with an asterisk. The vertical axis points towards North. The atmosphere parameters for each species are shown below each map. The column densities correspond to the values presented in \citeA{roth2021stable}. Note that in order to display the atmospheric structure of each species more clearly, the limits of the color bar are different between panels.}
\label{fig:atmospheric-models-R21}
\end{figure}

We now examine the case where $\mathrm{O}$ is present in the atmosphere model (i.e. $\mathrm{O_2} + \mathrm{O} + \mathrm{stable~H_2O}$), with column densities of the three species as derived by \citeA{roth2021stable}. In particular, $N_\mathrm{O}$ is set to $6\times10^{16}~\mathrm{m}^{-2}$ as the upper limit of its uncertainty range. Figure~\ref{fig:atmospheric-models-R21} shows an example of maps of column density in the trailing hemisphere for the three species of our atmospheric model, after integrating the neutral gas distributions along the line of sight with the subsolar point in the center of the disk. For this particular case, a cosine to the tenth-power on the angle to the subsolar point (i.e., $\beta=10$) is assumed for the $\mathrm{H_2O}$ distribution (panel (c)).

\begin{figure}[]
\centering
\noindent\includegraphics[width=1\textwidth]{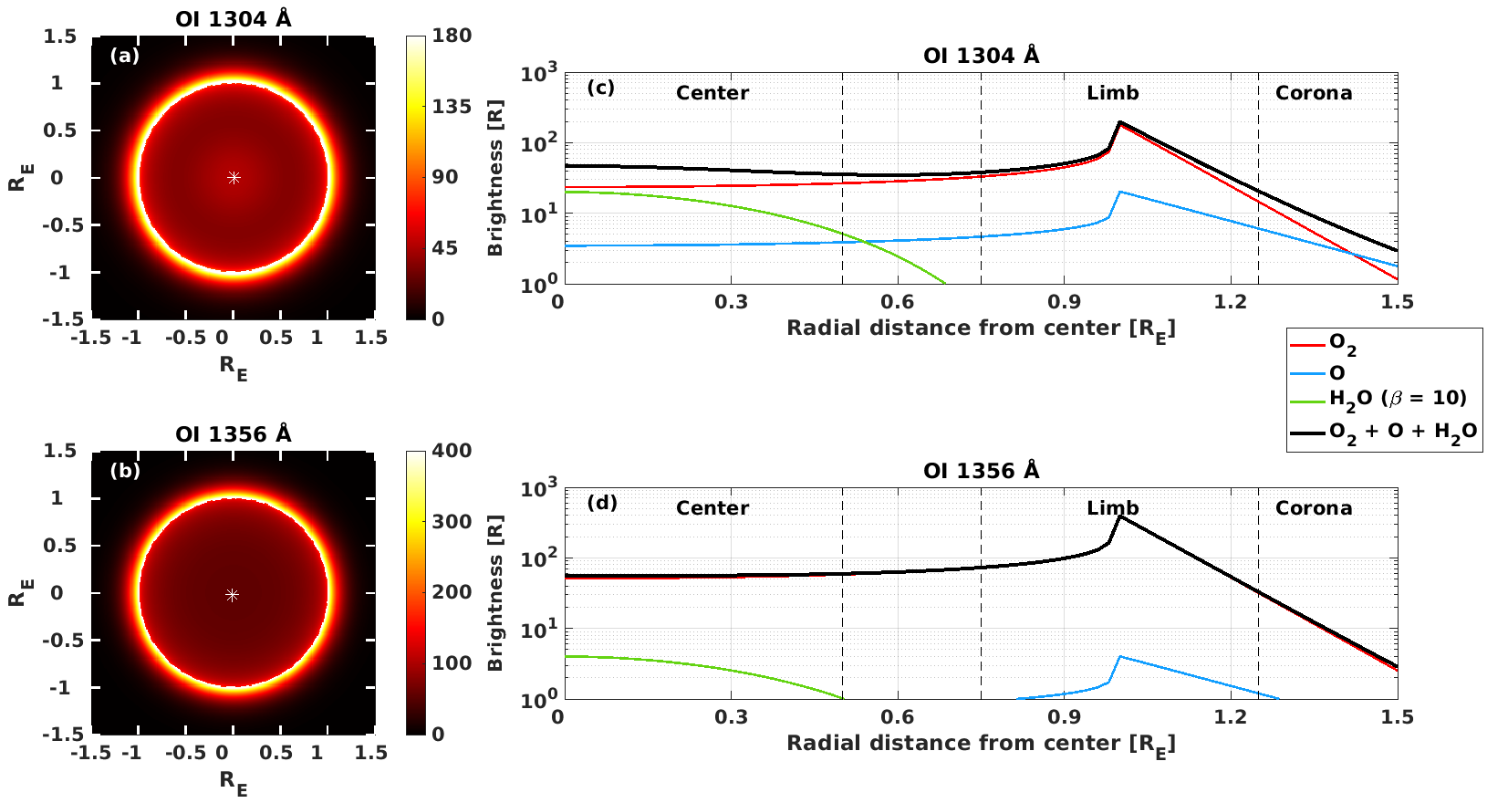}
\caption{(a) and (b) Images of the oxygen emission at 1304 \AA~and 1356 \AA~above Europa's trailing hemisphere for the total $\mathrm{O_2} + \mathrm{O} + \mathrm{H_2O}$ atmosphere model, whose individual components are presented in Figure~\ref{fig:atmospheric-models-R21}. The subsolar point is located at the center of the disk and is indicated with an asterisk. The vertical axis points towards North. (c) and (d) Radial  profiles of the average 1304 \AA~and 1356 \AA~brightness within concentric rings from the disk center out to $1.5~R_E$. The profiles are shown for the total atmosphere and for the individual contributions from $\mathrm{O_2}$, $\mathrm{O}$, and $\mathrm{H_2O}$.}
\label{fig:surface-brightness-radial-emissions-R21}
\end{figure}

Panels (a) and (b) of Figure~\ref{fig:surface-brightness-radial-emissions-R21} present the two-dimensional emission patterns of the total $\mathrm{O_2} + \mathrm{O} + \mathrm{stable~H_2O}$ atmosphere model whose individual components are shown in Figure~\ref{fig:atmospheric-models-R21}, for both OI 1304 \AA~and OI 1356 \AA~lines. These $361 \times 361$ pixel images, with a spacing of $0.01~R_E$, mainly reflect limb brightening from the global $\mathrm{O_2}$ and $\mathrm{O}$, with a minor contribution from $\mathrm{H_2O}$ to the total OI 1304 \AA~emission. The averaged simulated radial profiles in $0.025~R_E$ wide concentric rings (panels (c) and (d)) show contributions from the three neutral gases to the total OI 1304 \AA~brightness, with the emissions of $\mathrm{H_2O}$ being comparable to those of $\mathrm{O_2}$ close to the center of the disk. In contrast, at 1356 \AA, $\mathrm{H_2O}$ and $\mathrm{O}$ yield emissions which are more than one order of magnitude lower than those of $\mathrm{O_2}$, and thus the total averaged OI 1356 \AA~profile across the disk vastly originates from the latter.

\begin{figure}[]
\centering
\noindent\includegraphics[width=1\textwidth]{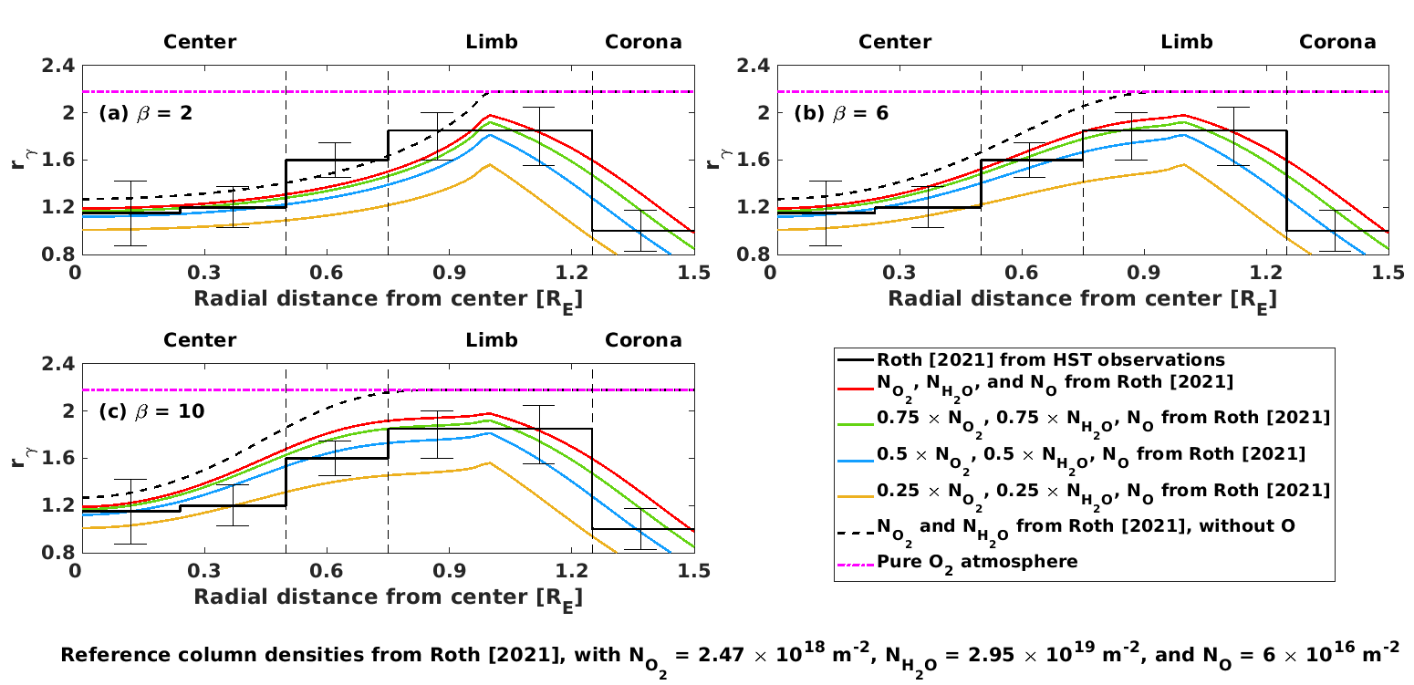}
\caption{Radial profile of the oxygen emission ratio of OI 1356 \AA~to OI 1304 \AA~for HST observations (black histogram) and for our simulated $\mathrm{O_2} + \mathrm{O} + \mathrm{H_2O}$ atmosphere models with varying $\mathrm{O_2}$ and $\mathrm{H_2O}$ column densities, as fractions from the values calculated by \citeA{roth2021stable}. Different panels correspond to varying degree of confinement of the stable $\mathrm{H_2O}$, as indicated by the exponent $\beta$.}
\label{fig:oxygen-emission-ratio-with-without-O}
\end{figure}

The observed and modelled oxygen emission ratio profiles are shown in panels (a) to (c) of Figure~\ref{fig:oxygen-emission-ratio-with-without-O} in solid black and red lines, respectively, for the same three values of $\beta$ (2, 6, and 10) examined in the previous section. The maximum of all profiles, both observed and simulated, is in the radial bins close to $1~R_E$, where the contributions from $\mathrm{O}$ and $\mathrm{H_2O}$ to the OI 1304 \AA~emissions are the lowest. At radial distances above the limb ($>1~R_E$), the abundance of $\mathrm{O}$ results in a higher 1304 \AA~than 1356 \AA~intensity, therefore reducing $r_\gamma$. The oxygen emission ratio also decreases towards the disk center, due to the increase in the $\mathrm{H_2O}$ column density and hence, in the emission due to $\mathrm{H_2O}$. It is worth emphasizing that \cite{roth2021stable} do not report any $\mathrm{H_2O}$ plumes active during the HST observations from which the oxygen emission ratio was derived, and therefore we do not take them into account in our simulated radial profile of $r_\gamma$. However, following the in-situ evidence provided by \cite{jia2018evidence} for the E12 flyby, we include a plume in the subsequent MHD modelling of the plasma interaction in the vicinity of the moon, as presented in Section~\ref{subsec:atmosphere model}.

The oxygen emission ratio derived from HST images by \citeA{roth2021stable} is provided with uncertainties along its radial profile, and therefore we seek further models that lie within the error bars of the observations. We calculate the emission ratio for three additional atmosphere models, in which we successively reduce the column densities of $\mathrm{O_2}$ and $\mathrm{H_2O}$ and multiply both original values in \citeA{roth2021stable} by 0.75, 0.5, and 0.25, while keeping the column density of $\mathrm{O}$ at the upper bound of
$6 \times 10^{16}~\mathrm{m}^{-2}$. The column densities of $\mathrm{O_2}$ and $\mathrm{H_2O}$ are decreased by the same percentage in each model, and therefore the abundance of $\mathrm{H_2O}$ with respect to $\mathrm{O_2}$ is 12 in all cases. Similar to the previous section, the scale heights of the species of our atmosphere model are kept constant, but the surface number density is recalculated in each case. The resulting radial profiles of the oxygen emission ratio are presented as solid green, blue, and yellow lines (for 75\%, 50\%, and 25\% of the $\mathrm{O_2}$ and $\mathrm{H_2O}$ column densities, respectively) in the three panels of Figure~\ref{fig:oxygen-emission-ratio-with-without-O}.

\subsubsection{Stable $\mathrm{H_2O}$ Distribution with $\beta = 2$}

The atmosphere model with $N_\mathrm{O_2}$, $N_\mathrm{H_2O}$, and $N_\mathrm{O}$ from \citeA{roth2021stable}, and with degree of confinement of the stable $\mathrm{H_2O}$ of $\beta = 2$, yields an oxygen emission ratio (red line in panel (a) of Figure~\ref{fig:oxygen-emission-ratio-with-without-O}
) that does not fit the observed profile in the bin around 0.6 $R_E$. Moreover, none of the simulated $r_\gamma$ profiles with decreased $\mathrm{O_2}$ and $\mathrm{H_2O}$ column densities, with $N_\mathrm{O}$ at its upper limit (in green, blue, and yellow), fit the observed emission ratio within its uncertainties. Therefore, we conclude that such an atmosphere model with a weakly concentrated stable $\mathrm{H_2O}$ around the subsolar point is not consistent with the HST data.

\subsubsection{Stable $\mathrm{H_2O}$ Distribution with $\beta = 6$}

Several of the atmospheric distributions assuming an $\mathrm{H_2O}$ component with $\beta = 6$ (panel (b)) fulfill the observed oxygen emission ratio within its uncertainties. The profile with the original column densities from \citeA{roth2021stable}, shown in red, fits the observations, except in the outermost bin, beyond 1.3 $R_E$. For comparison, the dashed black curve indicates the emission ratio for an atmosphere without atomic oxygen ($N_\mathrm{O} = 0$, as calculated in Section~\ref{subsec:oxygen emission ratio-no oxygen}). Therefore, the area between the solid red and dashed black profiles contains the emission ratios for all the models with fixed $N_\mathrm{O_2}$ and $N_\mathrm{H_2O}$ from \citeA{roth2021stable} and $\beta = 6$, but with varying $N_\mathrm{O}$ in the interval $0 \leq N_\mathrm{O} \leq 6\times10^{16}~\mathrm{m}^{-2}$. This means that models with $N_\mathrm{O} < 6\times10^{16}~\mathrm{m}^{-2}$, while keeping $N_\mathrm{O_2}$ and $N_\mathrm{H_2O}$ to the original values in \citeA{roth2021stable}, do not fit the observed $r_\gamma$ in the bin around 0.35 $R_E$.

It is also worth noting that the solid red line not only corresponds to the emission ratio of the model with the original $N_\mathrm{O_2},~N_\mathrm{H_2O},~\mathrm{and}~N_\mathrm{O}$ from \citeA{roth2021stable}, but also to the emission ratio of any model in which the column densities of the three species are simultaneously decreased by the same percentage. For instance, if all the abundances are reduced to 75\% of the values from \citeA{roth2021stable}, this can be expressed as follows:

\begin{equation}
r_{\gamma,75\%} = \frac{\mathrm{OI~1356~\AA}}{\mathrm{OI~1304~\AA}} = \frac{\sum_n\int_\mathrm{los} n_e \times 0.75n_n \times f_{n,\mathrm{1356~\AA}}~\mathrm{d}s}{\sum_n\int_\mathrm{los} n_e \times 0.75n_n \times f_{n,\mathrm{1304~\AA}}~\mathrm{d}s}  = r_{\gamma,100\%},
\label{eq:rgamma-ratio_withO}
\end{equation} 

\noindent where the subindex $n$ stands for each of the neutral species under consideration ($\mathrm{O_2}, \mathrm{H_2O}, \mathrm{and}~\mathrm{O}$). Analogous to equation~\ref{eq:rgamma-ratio}, this applies for all column densities, on the condition that the three of them are reduced by the same percentage with respect to the abundances in \citeA{roth2021stable}.

Decreasing the column density of $\mathrm{O_2}$ and $\mathrm{H_2O}$ to 75\%, while keeping the column density of $\mathrm{O}$ to its upper limit (green curve), yields a fit consistent with the observed $r_\gamma$. Moreover, all the profiles lying between the green and red ratios also fulfill the data within their error bars, i.e. any model with $N_\mathrm{O_2}$ and $N_\mathrm{H_2O}$ reduced to 75\% and with $N_\mathrm{O}$ ranging between 75\% and 100\% of the upper limit from \citeA{roth2021stable} explains the observed emission ratio within its uncertainties. Similarly, all the models with column densities of $\mathrm{O_2}$ and $\mathrm{H_2O}$ decreased by 50\% and with column density of $\mathrm{O}$ in the interval from 50\% to 100\% of the upper bound (between the red and blue lines, respectively), also fit the observed $r_\gamma$.

Finally, the profile in yellow, which belongs to the model with 25\% of the $N_\mathrm{O_2}$ and $N_\mathrm{H_2O}$ and 100\% of the $N_\mathrm{O}$ from \citeA{roth2021stable}, does not provide a good fit to the data. In contrast, reducing the column density of $\mathrm{O}$ to 25\% (red profile) is consistent with the observed $r_\gamma$. In addition, some of the models lying between the red and yellow curves also fit the emission ratio from \citeA{roth2021stable}, particularly those with $N_\mathrm{O}$ closer to 25\%.

\subsubsection{Stable $\mathrm{H_2O}$ Distribution with $\beta = 10$}
\label{subsubsec:beta10}

None of our atmosphere models with either 75\% or 100\% of the $\mathrm{O_2}$ and $\mathrm{H_2O}$ column densities from \citeA{roth2021stable}, and with $\beta = 10$, are consistent with the observed profile of $r_\gamma$, as the green and red lines in panel (c) of Figure~\ref{fig:oxygen-emission-ratio-with-without-O} show. This statement holds for all values of $N_\mathrm{O}$ between 0 and the upper limit of $6 \times 10^{16}~\mathrm{m}^{-2}$, as in all cases, the simulated $r_\gamma$ lies above the observations in the bins between the center and the limb. Hence, such a confined stable $\mathrm{H_2O}$ distribution invariably requires lower $\mathrm{O_2}$ and $\mathrm{H_2O}$ column densities, in order to agree with the emission ratio derived from the HST observations.

With $N_\mathrm{O_2}$ and $N_\mathrm{H_2O}$ as 50\% of the densities from \citeA{roth2021stable}, only the models with $N_\mathrm{O}$ close to 100\% of the upper limit fit the observed $r_\gamma$, considering the uncertainty range of the atomic oxygen abundance. Furthermore, none of the models with $N_\mathrm{O_2}$ and $N_\mathrm{H_2O}$ reduced to 25\% and with $N_\mathrm{O}$ equal to either 25\% or 100\% from \citeA{roth2021stable} (red and yellow profiles, respectively) are consistent with the observed $r_\gamma$. However, some of the atmosphere models with $\mathrm{O}$ column density ranging between both percentages are in agreement with the data.

In summary, our analysis of the oxygen emission ratio shows that the observations from \citeA{roth2021stable} only place conditional constraints on the column densities of $\mathrm{O_2}$ and $\mathrm{H_2O}$, given the uncertainty in the abundance of $\mathrm{O}$ and the spatial variability of the stable $\mathrm{H_2O}$. For each value of $\beta$, there exists a set of possible solutions that agree with the observed $r_\gamma$. With $\beta = 2$, only atmosphere models without $\mathrm{O}$ fit the data within its uncertainties between the center and the limb. In the case of a  moderately confined $\mathrm{H_2O}$ atmosphere ($\beta = 6$), models with 25\%, 50\%, 75\%, and 100\% of the $N_\mathrm{O_2}$ and $N_\mathrm{H_2O}$ from \citeA{roth2021stable} provide a good fit, conditioned on the abundance of atomic oxygen. Furthermore, the model with the most localized $\mathrm{H_2O}$ atmosphere ($\beta = 10$), requires $N_\mathrm{O_2}$ and $N_\mathrm{H_2O}$ lower than 75\% of the original values, and $N_\mathrm{O}$ close to its upper limit, in order to be consistent with the observed $r_\gamma$ profile. Since the parameter space of the column densities of $\mathrm{O_2}$ and $\mathrm{H_2O}$ is more constrained in the latter case, for the remainder of this work we mostly consider a strongly concentrated $\mathrm{H_2O}$ distribution around the subsolar point by fixing $\beta = 10$.

\begin{table}[]
\caption{$\mathrm{O_2}$ and $\mathrm{H_2O}$ column densities for different atmosphere models, as percentages (indicated in parenthesis) of the original values in \citeA{roth2021stable}. The last column provides the abundance of $\mathrm{H_2O}$ relative to $\mathrm{O_2}$.}
\centering
\begin{tabular}{c c c c}
\hline
Atmosphere model & $N_\mathrm{O_2}$ ($\times 10^{18}~\mathrm{m}^{-2}$) & $N_\mathrm{H_2O}$ ($\times 10^{19}~\mathrm{m}^{-2}$) & $N_\mathrm{H_2O}/N_\mathrm{O_2}$\\
 \hline
  1  & $2.47~(100\%)$ & $2.95~(100\%)$ & 12 \\
   2  & $1.85~(75\%)$  & $2.22~(75\%)$ & 12\\
   3  & $1.24~(50\%)$  & $1.47~(50\%)$ & 12\\
   4  & $0.62~(25\%)$  & $0.74~(25\%)$ & 12\\
   5  & $1.85~(75\%)$  & $2.95~(100\%)$ & 16\\
   6  & $1.24~(50\%)$  & $2.22~(75\%)$ & 18\\
 \hline
\end{tabular}
\label{tab:atmospheric models}
\end{table}

\begin{figure}[]
\centering
\noindent\includegraphics[width=1\textwidth]{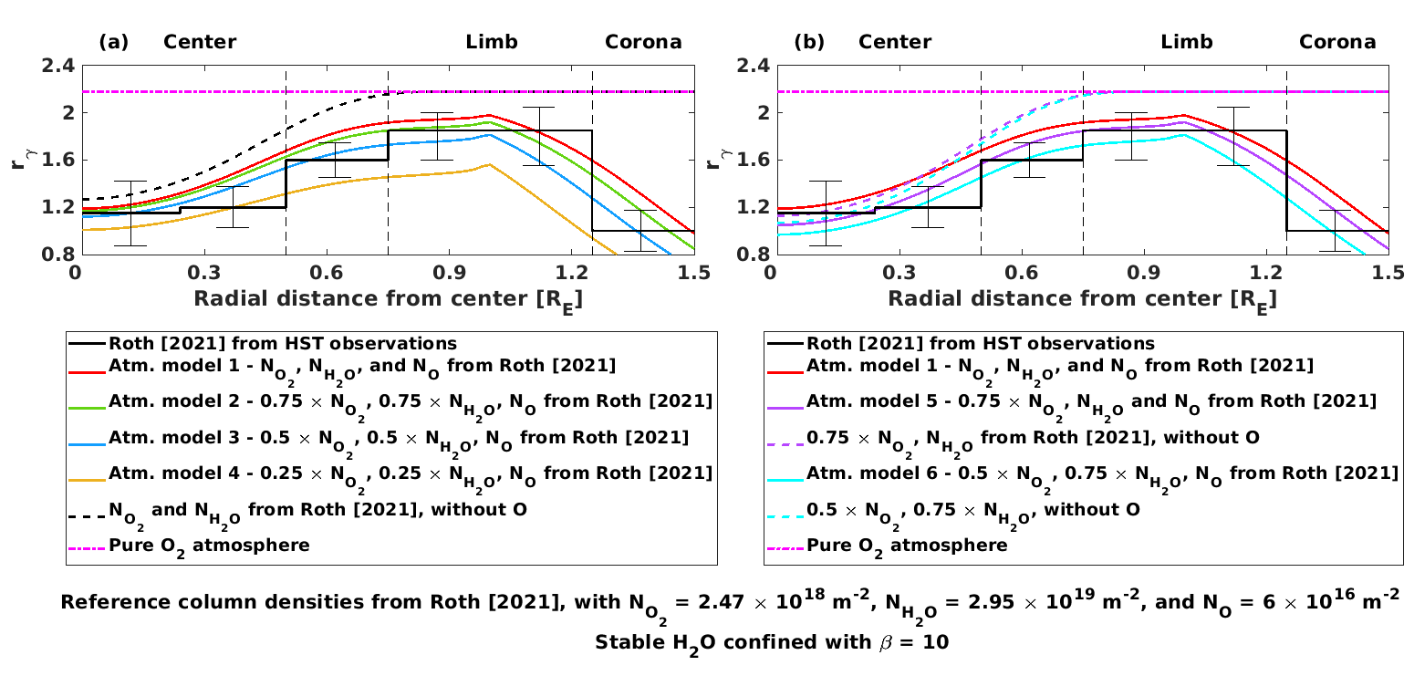}
\caption{Radial profile of the oxygen emission ratio of OI 1356 \AA~to OI 1304 \AA~for HST observations (black histogram) and for our simulated $\mathrm{O_2} + \mathrm{O} + \mathrm{H_2O}$ atmosphere models with varying $\mathrm{O_2}$ and $\mathrm{H_2O}$ column densities, as fractions of the values calculated by \citeA{roth2021stable}, and assuming $\beta = 10$. In panel (a) both column densities are decreased by the same rate, whereas in panel (b) the rate is different. This figure presents an overview of the models which will be employed in Section~\ref{sec:mhd results}, and it is consistent with the nomenclature of Table~\ref{tab:atmospheric models}.}
\label{fig:oxygen-emission-ratio-beta10}
\end{figure}

As presented in the previous paragraphs, the uncertainties along the observed $r_\gamma$ profile allow us to consider further several models by progressively decreasing $N_\mathrm{O_2}$ and $N_\mathrm{H_2O}$ from 100\% to 75\%, 50\%, and 25\% of the values from \citeA{roth2021stable}, while keeping the mixing ratio $N_\mathrm{H_2O}/N_\mathrm{O_2} = 12$. We summarize these distributions as models 1 to 4 in Table~\ref{tab:atmospheric models}. Panel (a) of Figure~\ref{fig:oxygen-emission-ratio-beta10} reiterates the simulated oxygen emission ratios for the particular case of $\beta = 10$, employing the nomenclature of Table~\ref{tab:atmospheric models}. These models will be employed in our MHD studies in Section~\ref{sec:mhd results} to further constrain Europa's neutral atmosphere.

Furthermore, on the basis of theoretical profiles of $r_\gamma$ as a function of the $\mathrm{O}/\mathrm{O_2}$ and $\mathrm{H_2O}/\mathrm{O_2}$ mixing ratios, and of the mean value $r_\gamma = 1.2$ at the center of the disk on the trailing side of HST spectral images, \citeA{roth2021stable} restricted the abundance of $\mathrm{H_2O}$ relative to $\mathrm{O_2}$ between 12 and 22. Therefore, the ratio $N_\mathrm{H_2O}/N_\mathrm{O_2} = 12$ of the models introduced previously is at the lower limit. We also examine models in which the column densities of $\mathrm{O_2}$ and $\mathrm{H_2O}$ are decreased with respect to the original values in \citeA{roth2021stable} by different percentages. We find two combinations of column densities that yield ratios $N_\mathrm{H_2O}/N_\mathrm{O_2}$ within 12 to 22, and these are included as models 5 and 6 in Table~\ref{tab:atmospheric models}. 

We first calculate the oxygen emission ratio while keeping the column density of $\mathrm{O}$ at its upper bound. The resulting profiles for these two atmospheric distributions are depicted by the solid purple and cyan lines in panel (b) of Figure~\ref{fig:oxygen-emission-ratio-beta10}, respectively. Since in these two models the abundance of $\mathrm{H_2O}$ is decreased by a smaller percentage than $\mathrm{O_2}$, the simulated emission ratio is displaced below the observed profile, reaching values close to 1 in the center of the disk, as this is the location in which we concentrate our $\mathrm{H_2O}$ atmosphere (i.e., the subsolar point). We also calculate the emissions for models 5 and 6 neglecting atomic oxygen, and the corresponding $r_\gamma$ is depicted by the dashed curves in panel (b). Given the uncertainty range of the $N_\mathrm{O}$, our results suggest that 75\% of the $N_\mathrm{O_2}$ in combination with 100\% of the $N_\mathrm{H_2O}$ from \citeA{roth2021stable} yields an emission ratio consistent with the data. Lower values of $N_\mathrm{O_2}$ and $N_\mathrm{H_2O}$, such as in model 6, also fit the observed profile within its uncertainties.

With the results presented in this section, we identify several $\mathrm{O_2}$ and $\mathrm{H_2O}$ column densities that fulfill the observed oxygen emission ratio within its error bars. However, due to the uncertainty in the $\mathrm{O}$ column density and the spatial extent of the stable $\mathrm{H_2O}$, the observed $r_\gamma$ profile from \citeA{roth2021stable} only conditionally restricts the abundances of $\mathrm{O_2}$ and $\mathrm{H_2O}$. The MHD simulations presented in the next section will provide additional information to constrain the column densities of these species in Europa's neutral atmosphere. 

\section{MHD Simulations of the Galileo E12 Flyby}
\label{sec:mhd results}

\begin{figure}
\centering
\noindent\includegraphics[width=0.7\textwidth]{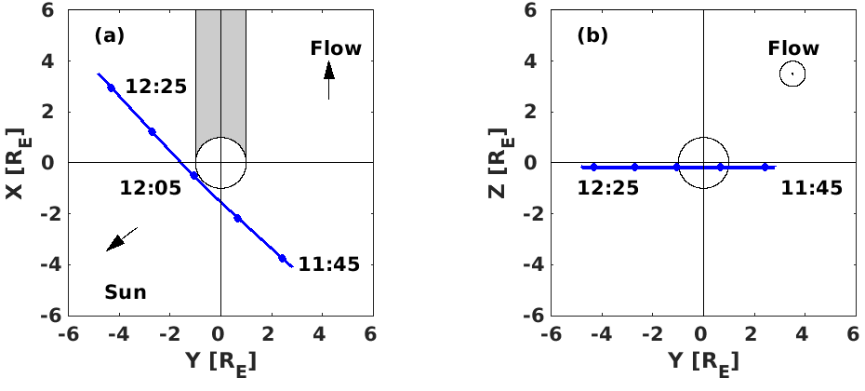}
\caption{E12 flyby trajectory in the (a) XY and (b) YZ planes. The gray shaded region in panel (a) indicates Europa's downstream geometric wake.}
\label{fig:flyby geometry}
\end{figure}

We now quantitatively investigate Europa's interaction with its plasma environment for the conditions of the Galileo E12 flyby by means of the MHD model, as described in Section \ref{sec:mhd}, and we also compare our simulations with the magnetic field measurements collected by the magnetometer. Out of the eight targeted passes in which MAG data was acquired, this flyby came closest to the surface (196 km). In addition, this pass crossed the trailing sunlit hemisphere of Europa near the equator ($-8.6\degree$), where the abundance of the stable $\mathrm{H_2O}$ detected by \citeA{roth2021stable} is expected to be maximum. The geometry of this flyby, illustrated in Figure \ref{fig:flyby geometry}, makes it ideal to test our candidate atmospheric models, and in particular, to elucidate the contribution of $\mathrm{H_2O}$ located around the subsolar point to the plasma interaction. The E12 pass occurred on 16 December 1997 and remained below 400 km altitude between 12:00:59 and 12:05:37 UT, with closest approach to Europa's surface at 12:03:20 UT \cite{kivelson1999europa, jia2018evidence}. In addition, Galileo's trajectory was the closest to the subsolar point at 12:03:54 UT.  The spacecraft traversed upstream in the plasma flow at the center of the plasma torus, with magnetic latitude relative to Jupiter's magnetic equator of $-0.8\degree$. Europa's system III longitude at the time of the flyby was $118\degree$. 

\begin{figure}[]
\centering
\noindent\includegraphics[width=1\textwidth]{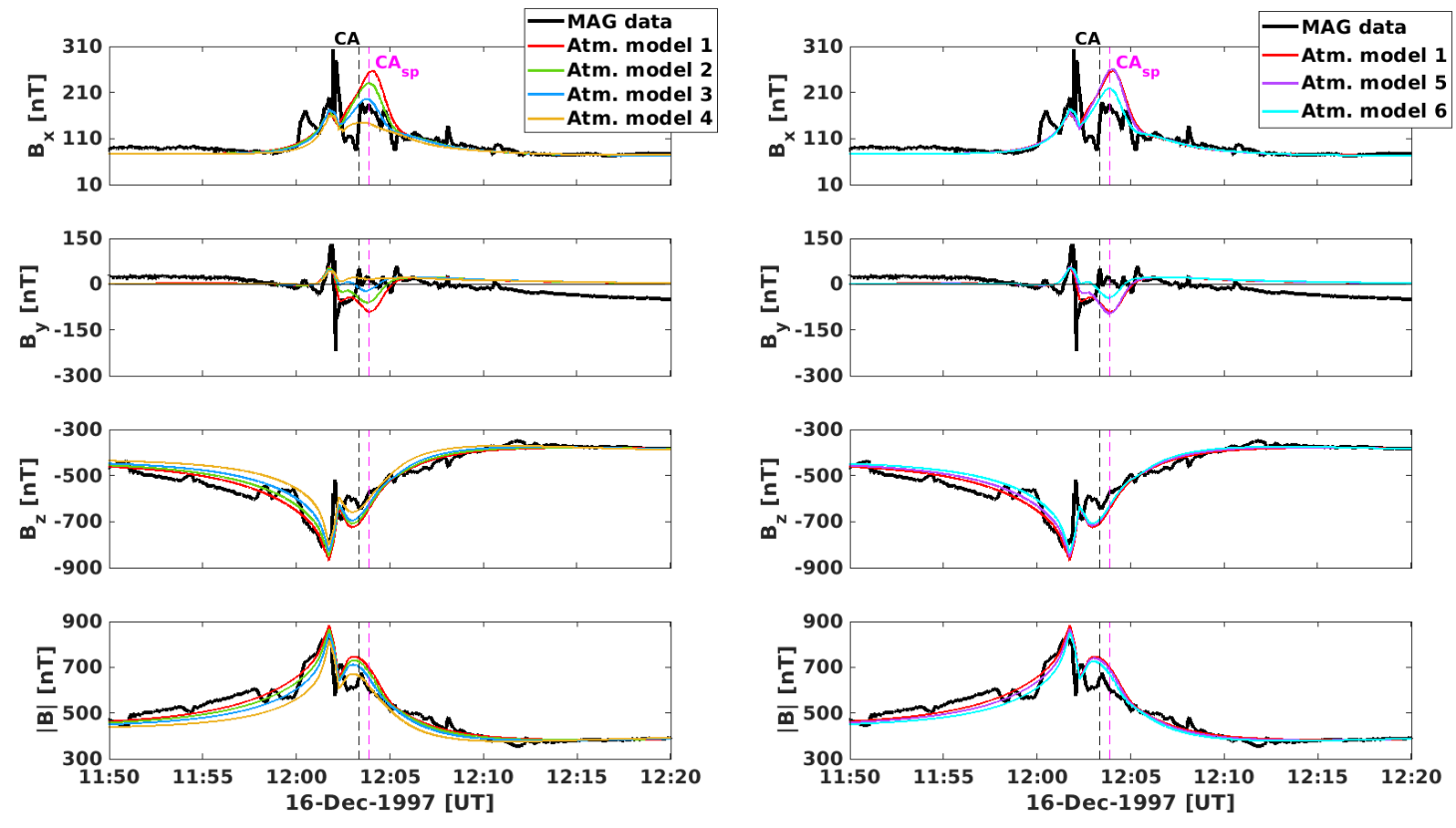}
\caption{Galileo E12 flyby. Black lines show MAG data. Color coded are different simulations with varying 
$\mathrm{O_2}$ and $\mathrm{H_2O}$ densities.
In the left column, values are reduced by the same percentage of $\mathrm{O_2}$ and $\mathrm{H_2O}$, whereas in the right, the mixing ratio of $\mathrm{O_2}$ and $\mathrm{H_2O}$ has been changed. Properties of the atmospheric models are listed in Table \ref{tab:atmospheric models}. The vertical dashed black and magenta lines indicate Galileo's closest approach to Europa's surface (CA) and Galileo's closest approach to the subsolar point ($\mathrm{CA_{sp}}$), respectively.}
\label{fig:zeus-R21-75-50}
\end{figure}

The magnetometer data for E12 flyby is shown in Figure \ref{fig:zeus-R21-75-50}. The magnetic field was unusually large upstream of Europa. From $\sim$12:00 UT to $\sim$12:03 UT, all three components of the magnetic field fluctuated. The sudden perturbations by hundreds of nT about one minute before closest approach were attributed by \citeA{jia2018evidence} to the passage of Galileo through a water plume. 

In this study we aim to answer if the perturbation after closest approach, between $\sim$12:03:30 UT and $\sim$12:05:30 UT, mainly evident as a local maximum in the $B_x$ component, is imposed by the presence of a stable $\mathrm{H_2O}$ atmosphere located at the subsolar point. In order to test this hypothesis we conduct several MHD simulations, in which we assume a neutral atmosphere consisting of global $\mathrm{O_2}$, $\mathrm{H_2O}$ localized at the subsolar point (with $\beta = 10$), and a plume (as described in Section \ref{subsec:atmosphere model}). The column densities of the first two components are varied according to the atmospheric models 1 to 6 presented in Table~\ref{tab:atmospheric models} of the previous section, whereas the properties of the plume are kept constant in all simulations. 

Figure \ref{fig:zeus-R21-75-50} compares the magnetic field measured by Galileo with the model results extracted from the simulations along the spacecraft trajectory. The left column shows the cases in which the densities of $\mathrm{O_2}$ and $\mathrm{H_2O}$ at the subsolar point are kept as in \citeA{roth2021stable} or both are decreased keeping the mixing ratio constant (atmospheric models 1 to 4). The panels on the right hand side of  Figure \ref{fig:zeus-R21-75-50} present the results for the models in which the mixing ratios are not constant (models 5 and 6). 

In the dense $\mathrm{H_2O}$ atmosphere confined around the subsolar point, electron impact ionization and ion-neutral collisions are enhanced, and therefore, stronger magnetic field perturbations are generated. In all our simulations, perturbations are observed after closest approach, in accordance with MAG data. In particular, the largest modelled perturbation in the $x$ component takes place around Galileo's closest approach to the subsolar point, where \citeA{roth2021stable} suggested the stable $\mathrm{H_2O}$ distribution to be maximum. However, the predicted $B_x$ fluctuations are largely overestimated ($> 40~\mathrm{nT}$) by atmospheric models 1, 2, and 5, namely those with both $\mathrm{O_2}$ and $\mathrm{H_2O}$ column densities $\geq 75\%$ from the values derived by \citeA{roth2021stable}. Model 3, with densities reduced by $50\%$, provides the best agreement to the perturbations after closest approach, deviating $13~\mathrm{nT}$ at the location of the local maximum in $B_x$. Model 6 yields the second best fit to the data and predicts variations with amplitudes between those of models 2 and 3.

On the contrary, model 4, with $25\%$ of the $\mathrm{O_2}$ and $\mathrm{H_2O}$ abundances from \citeA{roth2021stable}, provides a poor fit to the magnetic field data, as it underestimates the observed perturbation in $B_x$ by 50 nT. For this particular atmospheric model, our analysis of $r_\gamma$ in Section~\ref{subsubsec:beta10} showed that $\mathrm{O}$ column densities 25\% to 100\% of the upper limit from \citeA{roth2021stable} would yield profiles consistent with the observed ratio within its error bars. However, the lack of agreement between the magnetic field measurements and the MHD simulations effectively rules out model 4 as a candidate for Europa's atmosphere. This highlights the fact that a joint analysis of HST spectral images and Galileo MAG data is important to generate new constraints on the composition of Europa's neutral environment.

Similarly to $B_x$, the $B_y$ component is best reproduced with model 3, followed by model 6, whereas the remainder do not provide a satisfactory fit to the measurements. $B_z$ produces similar magnetic field responses in all cases. Therefore, we deem $B_x$ as the most diagnostic component to identify model 3 as the best out of the six candidates. As the parameters of the plume (column density, location, and tilt) do not vary between simulations, the abrupt large-amplitude fluctuations linked to this feature are similar in all cases. Our MHD simulations allow us to conclude that 50\% of the $N_\mathrm{O_2}$ and between 50\% and 75\% of the $N_\mathrm{H_2O}$ from \citeA{roth2021stable} (i.e. models 3 and 6) are required to reproduce the amplitude and location of the magnetic field perturbations after closest approach. These findings are consistent with the analysis of the oxygen emission ratio, presented in Section~\ref{sec:oxygen emission ratio}, provided that $N_\mathrm{O}$ is close to the upper limit derived by \citeA{roth2021stable}.

\begin{figure}[]
\centering
\noindent\includegraphics[width=1\textwidth]{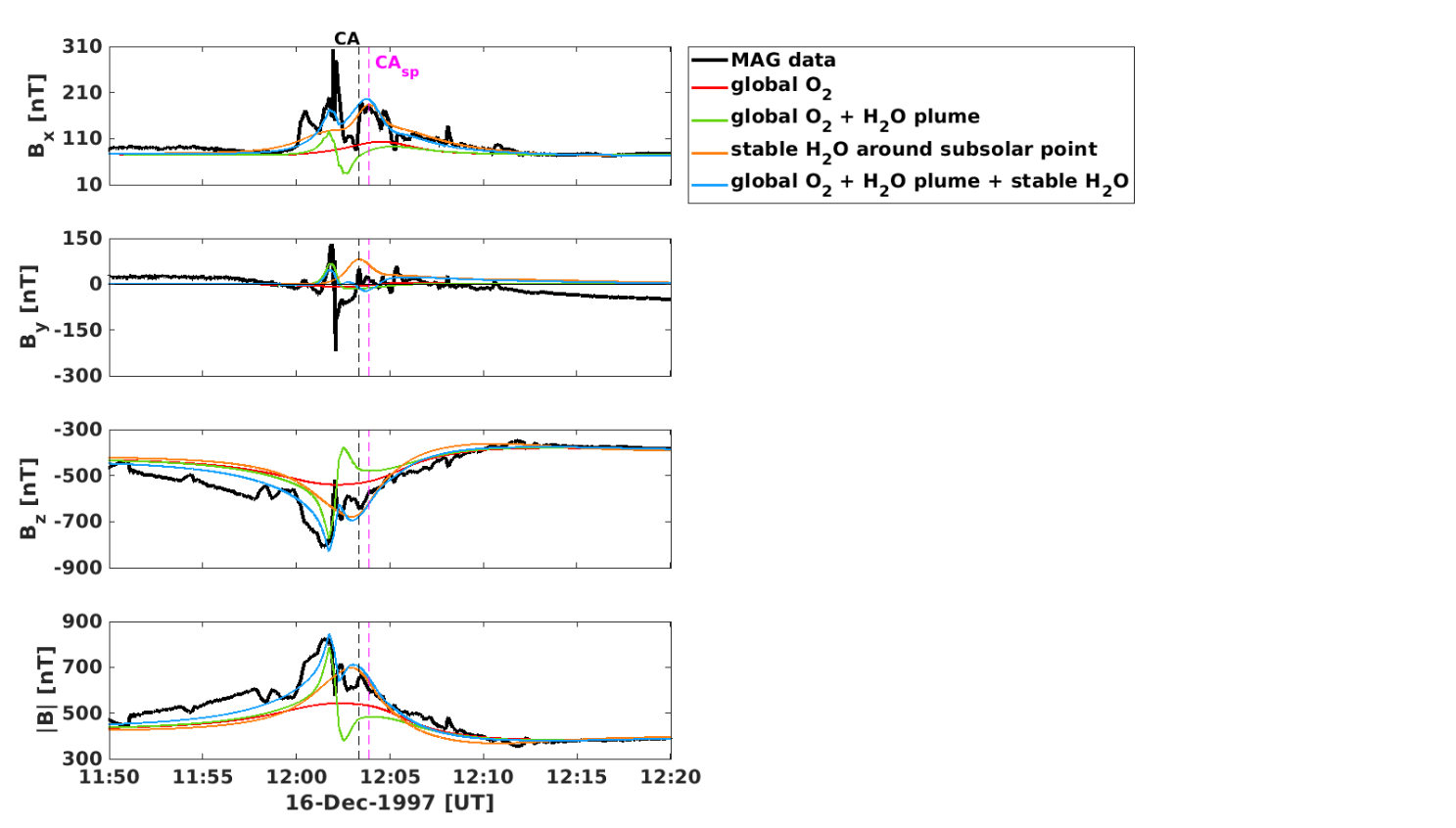}
\caption{Contributions of $\mathrm{O_2}$ and $\mathrm{H_2O}$ atmospheric components. The black line indicates Galileo MAG data; and the red, green, orange, and blue lines give the simulated magnetic fields for the E12 flyby trajectory assuming different atmospheric constituents. The column densities of $\mathrm{O_2}$ and $\mathrm{H_2O}$ at the subsolar point correspond to atmospheric model 3.}
\label{fig:zeus-R21-50-components}
\end{figure}

In Europa's atmosphere, molecular $\mathrm{O_2}$ is distributed approximately uniformly around the moon, whereas $\mathrm{H_2O}$ is present as a confined component, either in the form of sporadic plumes, a stable concentration around the subsolar point, or the combination of both. In order to better understand the effects of the individual contributions of each species on the plasma interaction, we perform further MHD simulations with the atmospheric model that overall provided the best fit to the HST and MAG data, i.e. model 3, by successively adding, one at a time, each element of our three-component atmosphere (Figure \ref{fig:zeus-R21-50-components}).
We start by considering only a global radially symmetric $\mathrm{O_2}$ distribution. Since our $\mathrm{O_2}$ column density is in the lower end of the typical range between 2 and $15 \times 10^{18}~\mathrm{m}^{-2}$ \cite{hall1995detection, hall1998far,mcgrath2009observations, roth2014transient} there is very little contribution from this species to the plasma interaction. The variations around closest approach are of low amplitude, $\sim$30 nT and $\sim$100 nT in $x$ and $z$ respectively, relative to the background values. The addition of the water plume to our model predicts the abrupt and rapid fluctuation of magnetic field prior to closest approach, similar to the simulations in \citeA{jia2018evidence}. However, the variations between $\sim$12:03 UT and $\sim$12:06 UT are not reproduced by the model, as can be seen in e.g. $B_x$ and the total field $|B|$. 

The individual contribution of the stable $\mathrm{H_2O}$ atmosphere centered at the subsolar point is mainly evident as a local maximum in $B_x$, where the magnetic field is enhanced by 80 nT just after closest approach. The perturbation in the modelled $x$ component is concurrent with the observed fluctuations, and both lie within the region where the $\mathrm{H_2O}$ distribution is predicted to be the most abundant, i.e., the subsolar point. Our $\mathrm{H_2O}$ atmosphere also reproduces some of the variations observed in $B_y$ and the gradual recovery of $B_z$ after closest approach.  

Lastly, when all three atmospheric constituents are taken into account, two substructures of a confined nature are evident in $B_x$. Such features cannot be produced by a globally distributed $\mathrm{O_2}$, and are therefore indicative of a localized component, as is the case of water. Between the occurrence of the plume and the stable $\mathrm{H_2O}$, just before closest approach, the measured $B_x$ and $B_y$ components decrease abruptly, while $B_z$ is enhanced. However, our simulations do not reproduce such variations as markedly as the MAG data show. We interpret this lack of agreement as a consequence of our parametrization of the water plume, which does not fully resolve the perturbed magnetic field nor the sharpness of the gradients adjacent to this structure. Nevertheless, the focus of our study is after the time of closest approach, when the signature of $\mathrm{H_2O}$ centered around the subsolar point is present in the data and our simulations.

\section{Discussion}

Our MHD simulations demonstrate that $N_\mathrm{O_2}$ has to be reduced to $50\%$ and $N_\mathrm{H_2O}$ between $50\%$ and $75\%$ with respect to the values in \citeA{roth2021stable}, and lie within the error bars of the observed oxygen emission ratio, in order to fulfill the conditions posed by HST and MAG data. In this section we assess the robustness of this finding by varying certain parameters of the atmospheric and MHD models ($\mathrm{H_2O}$ distribution and electron impact ionization, respectively) and performing three sets of additional simulations. We specifically consider the atmospheric distribution described by model 3, i.e. $\mathrm{O_2}$ and $\mathrm{H_2O}$ abundances decreased to 50\% of the values from \citeA{roth2021stable}, and $\mathrm{O}$ column density at the upper limit.

\subsection{Degree of Confinement of the Stable $\mathrm{H_2O}$}

At first, we vary the spatial extent of $\mathrm{H_2O}$ around the subsolar point, as described by the exponent $\beta$ of the cosine term in equation (\ref{eq:h2o density}). The MHD simulations presented previously employed $\beta = 10$, and we examine four additional cases: $\beta = 2, 4, 6,$ and 8. As presented in Section~\ref{sec:oxygen emission ratio}, the rate of decrease of the $\mathrm{H_2O}$ column density with respect to the distance from the subsolar point is faster as $\beta$ increases.

\begin{figure}[]
\centering
\noindent\includegraphics[width=1.1\textwidth]{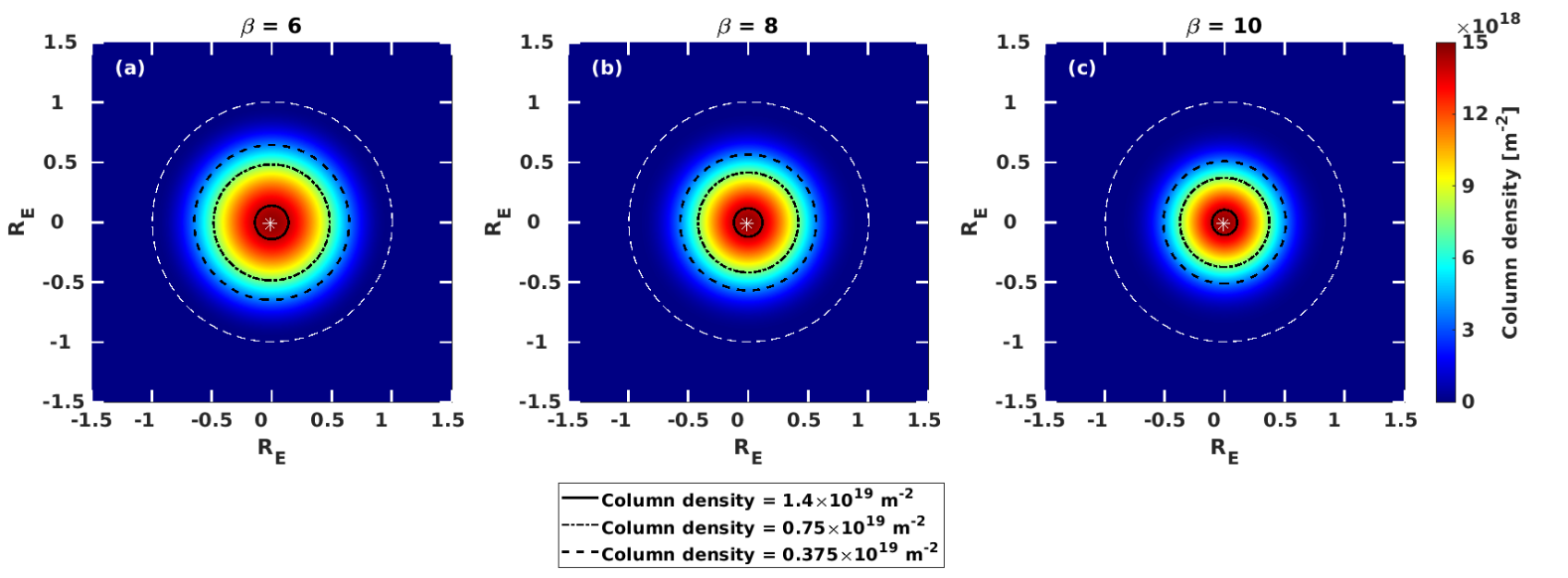}
\caption{Degree of confinement of the $\mathrm{H_2O}$ component. Line-of-sight integrated column density maps in the trailing hemisphere for the indicated values of the exponent $\beta$. The subsolar point is located at the center of the disk and is indicated with an asterisk. The vertical axis points towards North. Black contours indicate column densities of $95\%$ (solid), $50\%$ (dashed-dotted), and $25\%$ (dotted) of the maximum at the disk center, corresponding to 1.4, 0.75, and $0.375 \times 10^{19}~\mathrm{m}^{-2}$, respectively.}
\label{fig:atmospheric-models-R21-50-beta}
\end{figure} 

The line-of-sight integrated column density of the stable $\mathrm{H_2O}$ component with the three most confined distributions ($\beta = 6, 8$, and 10) is illustrated in Figure \ref{fig:atmospheric-models-R21-50-beta}. For ease of comparison among the three cases, contours corresponding to column densities of 1.4, 0.75, and $0.375 \times 10^{19}~\mathrm{m}^{-2}$ are overlaid. These values indicate $95\%$, $50\%$, and $25\%$ of the maximum abundance at the subsolar point, respectively.  As expected, the radial extent of the $\mathrm{H_2O}$ distribution is more confined with increasing $\beta$. For the exponent $\beta = 6$, the contour with column density of $1.4 \times 10^{19}~\mathrm{m}^{-2}$ is located at $0.18~R_E$ from the center, whereas for $\beta = 10$, this contour is found at $0.1~R_E$. For the column density equal to  $0.375 \times 10^{19}~\mathrm{m}^{-2}$, the contours extend from $0.51~R_E$ ($\beta = 10$) to $0.67~R_E$ ($\beta = 6$).

\begin{figure}[]
\centering
\noindent\includegraphics[width=1\textwidth]{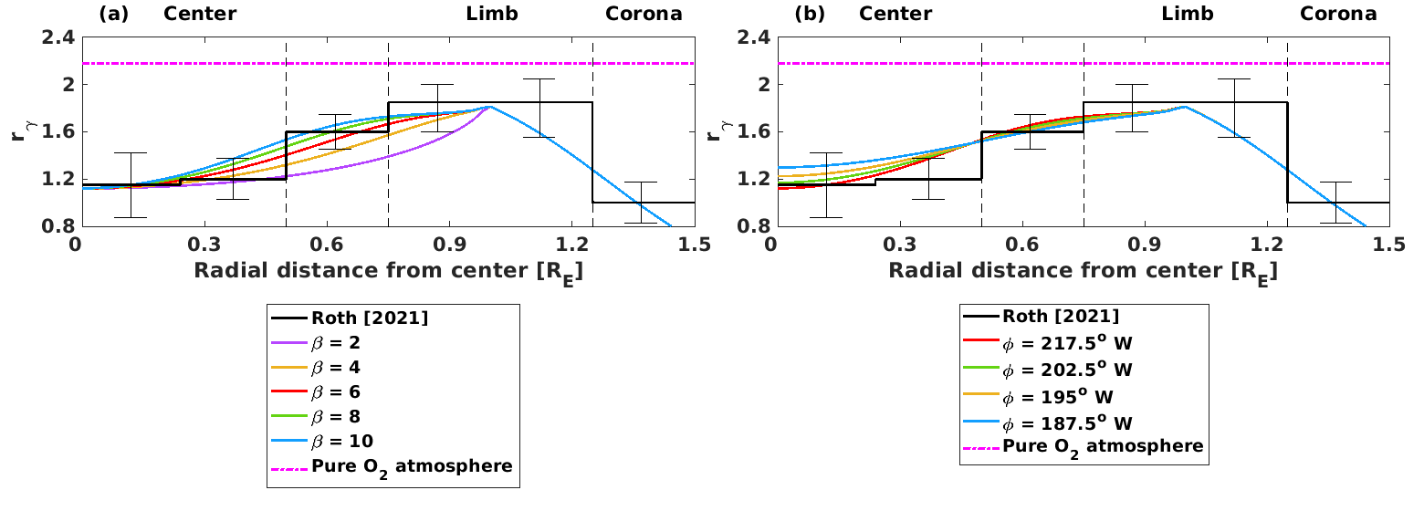}
\caption{Sensitivity analysis of the stable $\mathrm{H_2O}$ atmosphere: degree of confinement and location of maximum. Radial profiles of the observed and simulated oxygen emission ratio for our $\mathrm{O_2} + \mathrm{O} + \mathrm{H_2O}$ atmosphere model with $\mathrm{O_2}$ and $\mathrm{H_2O}$ column densities $50\%$ from the values in \citeA{roth2021stable}. Panel (a) presents the results for different values of the exponent $\beta$ in the $\mathrm{H_2O}$ distribution, and panel (b) for various locations of the center of the $\mathrm{H_2O}$ component (longitude $\phi$).}
\label{fig:oxygen-emission-ratio-R21-50-beta-subsolarpoint}
\end{figure}

We start by calculating the oxygen emission ratio for the total $\mathrm{O_2} + \mathrm{O} + \mathrm{stable~H_2O}$ atmosphere, with the column densities belonging to model 3. Panel (a) of Figure \ref{fig:oxygen-emission-ratio-R21-50-beta-subsolarpoint} shows the profiles for the five cases of $\beta$. The least confined distribution, with $\beta = 2$, does not match the observed $r_\gamma$ profile beyond $\sim$0.5 $R_E$. The remainder of the exponents provide a satisfactory fit to the HST observations within the error bars across the entire disk, and they yield similar values of $r_\gamma$ at the central bins. Nonetheless, the profiles diverge the most between 0.3 and 0.8 $R_E$, where the $\mathrm{H_2O}$ in the model with $\beta = 4$ is the least confined, and thus $r_\gamma$ is the lowest. 

\begin{figure}
\centering
\noindent\includegraphics[width=1\textwidth]{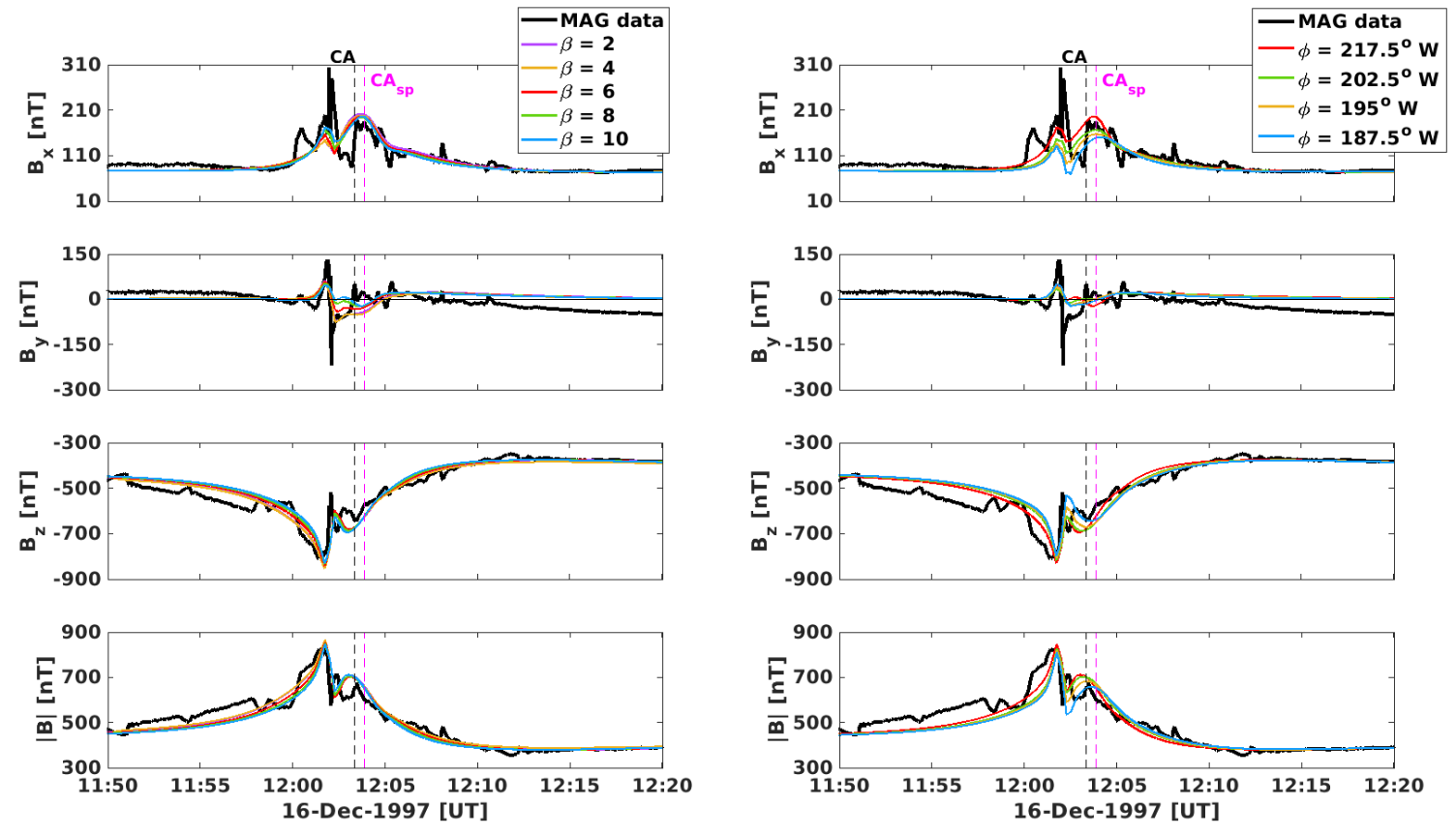}
\caption{Sensitivity analysis of the stable $\mathrm{H_2O}$ atmosphere: degree of confinement and location of maximum. The black line indicates Galileo MAG data for the E12 flyby trajectory. Color coded are different simulated magnetic fields for various values of $\beta$ in the $\mathrm{H_2O}$ distribution (left column) and locations of the center
of the $\mathrm{H_2O}$ component (longitude $\phi$, right column).}
\label{fig:zeus-R21-50-beta-subsolarpoint}
\end{figure}

After assessing to what extent these distributions are consistent with the HST observations, we use them to conduct MHD simulations of the plasma interaction (Figure \ref{fig:zeus-R21-50-beta-subsolarpoint}, left column). It must be emphasized that the only parameter that differs among them is the exponent $\beta$. In all cases, the simulated magnetic field at the time of closest approach to the subsolar point is comparable in amplitude, as this is the location at which the stable $\mathrm{H_2O}$ column density reaches its maximum for the five $\beta$ values. The field magnitude for $\beta = 2$ is only marginally larger by $\sim$20 nT with respect to the model with $\beta = 10$, since an $\mathrm{H_2O}$ atmosphere with lower $\beta$ is spatially wider, contains more neutrals available for collisions, and therefore generates larger magnetic field perturbations. We have also explored other functional forms describing narrower $\mathrm{H_2O}$ distributions, e.g., exponential or trigonometric multiplied by a scalar, but the resulting oxygen emission ratio diverges from the observed profile at the center of the disk and does not fit it within its uncertainties. All in all, the similarity among our five simulations in the left panel of Figure \ref{fig:zeus-R21-50-beta-subsolarpoint} shows that the exact value of $\beta$ and the spatial extent of the stable $\mathrm{H_2O}$ cannot be uniquely constrained with the MAG data, thereby highlighting the importance of simultaneously exploring the structure and density of the atmosphere with the HST spectral images.

\subsection{Location of the Stable $\mathrm{H_2O}$}

A second parameter that we vary is the location of the center of the stable $\mathrm{H_2O}$ component. Our previous simulations assume that the maximum $\mathrm{H_2O}$ abundance is aligned with the instantaneous subsolar point. However, thermal inertia of Europa's icy surface might shift the location with the largest temperature, and thus of the maximum $\mathrm{H_2O}$ density, with respect to the subsolar point. In this regard, the brightness temperature profiles presented by \citeA{spencer1999temperatures} suggest a thermal delay relative to the subsolar point. To investigate this, we displace the center of the $\mathrm{H_2O}$ distribution in longitude from $\phi = 217.5\degree$ W (corresponding to 12 LT), towards the east (in the afternoon sector), by $15 \degree$, $22.5 \degree$, and $30 \degree$. As in the previous case, we first make certain that these models are consistent with the HST data by calculating the oxygen emission ratio (panel (b) of Figure \ref{fig:oxygen-emission-ratio-R21-50-beta-subsolarpoint}). The four profiles fit the observed $r_\gamma$ within its uncertainties in all the bins except between 0.25 and 0.5 $R_E$, where the modelled values for the cases with $\phi = 195 \degree$ W and $\phi = 187.5 \degree$ W (corresponding to 1.5 and 2 hours after 12 LT) fall out of the error bars by 0.75 of $r_\gamma$. In the central bin, between 0 and $0.25~R_E$, and for the atmosphere with $\mathrm{H_2O}$ coincident with the subsolar point at 12 LT (in red), $r_\gamma$ is the lowest. For the model with the most displaced $\mathrm{H_2O}$ distribution (in blue), the $\mathrm{H_2O}$ density at the center of the disk is lower, $\mathrm{O_2}$ dominates, and thus $r_\gamma$ is larger by 0.17. The opposite pattern is observed at the limb, between 0.6 and 0.9 $R_E$, where the profile for the non-displaced subsolar point is larger by 0.07 relative to the most displaced one. The location at which this trend reverses is $\sim$0.48 $R_E$. 

The panels on the right hand side of Figure \ref{fig:zeus-R21-50-beta-subsolarpoint} compare the magnetic field measured by MAG and the predicted field with different locations of the maximum of the stable $\mathrm{H_2O}$ compared to the subsolar point. The remainder of the parameters stay unchanged between simulations. The four cases reproduce the local maximum in $B_x$ after closest approach, consistent with the presence of $\mathrm{H_2O}$ at this location. As already mentioned, our initial simulation with the $\mathrm{H_2O}$ abundance centered at the subsolar point overestimates the observed field by $\sim$13 nT in the $x$ component. On the contrary, the other three simulations, with the displaced stable $\mathrm{H_2O}$ distribution, underestimate the measured $B_x$ by $\sim$16 nT ($\phi = 202.5\degree$ W) and $\sim$32 nT ($\phi = 187.5\degree$ W). The occurrence of the local maximum is also displaced from 12:03:46 UT for $\phi = 217.5\degree$ W to 12:04:10 UT for $\phi = 187.5\degree$ W. The other two components are also reproduced similarly with the different $\mathrm{H_2O}$ models. The field magnitude $|B|$ decreases abruptly after the peak due to the plume (at 12:01:40 UT) by 207 nT and 271 nT in the simulations with the maximum $\mathrm{H_2O}$ at 12 LT and 2 hours after, respectively. Both values are in accordance with the observed decrease of 248 nT. In analogy to $B_x$, the local maximum in $|B|$ around closest approach occurs the earliest in the simulation with the stable $\mathrm{H_2O}$ centered at the subsolar point. Our findings show that the plasma interaction is sensitive to the location of the $\mathrm{H_2O}$ atmosphere, whose center might be misaligned with respect to the subsolar point.

The simulations with $\phi = 217.5\degree$ W and $\phi = 202.5\degree$ W are the best constrained both by HST and MAG measurements. For the latter case, the location of the $\mathrm{H_2O}$ maximum is displaced one hour after 12 LT. Therefore, our results suggest that the plasma interaction for the $\mathrm{H_2O}$ atmosphere is partly dictated by Europa's surface temperature, which in turn controls the sputtering and sublimation yield of water ice \cite{fama2008sputtering, plainaki2013exospheric, vorburger2018europa}. These findings also hint that thermal inertia might play a role in the location of the $\mathrm{H_2O}$ atmosphere. 

The stable $\mathrm{H_2O}$ distribution is concentrated in the vicinity of the subsolar point, but its column density is too large as expected from standard temperature maps of Europa's surface. An $\mathrm{H_2O}$ column density of $N_\mathrm{H_2O} = 1.47 \times 10^{19}~\mathrm{m}^{-2}$ would require a temperature of 142 K, in contrast to the observed maximum dayside value of 132 K \cite{spencer1999temperatures}. The modelling works of \citeA{smyth2006europa} and \citeA{vorburger2018europa} have considered both sublimation and sputtering as sources. Assuming surface temperatures between 95 and 132 K, their predicted $\mathrm{H_2O}$ column densities lie between 2 and $6 \times 10^{16}~\mathrm{m}^{-2}$. Therefore, an additional mechanism is required to explain this density surplus. \citeA{roth2021stable} speculated that sputtering and secondary sublimation might be the origin of the detected stable $\mathrm{H_2O}$ atmosphere, in line with the results of \citeA{teolis2017plume}. In addition, \citeA{addison2022effect} also observed an increase in the $\mathrm{H_2O}$ sputtering rate and a preferential emission of $\mathrm{H_2O}$ molecules near the upstream apex, in agreement with the findings of \citeA{roth2021stable}. 

\subsection{Electron Impact Ionization Rate}

As pointed out by \citeA{roth2021stable}, the derived abundances of $\mathrm{O_2}$, $\mathrm{O}$, and $\mathrm{H_2O}$ in their model are sensitive to the assumed electron properties, i.e., density and temperature. The electron impact ionization rate $f_\mathrm{imp}$ depends on the density of neutrals and electrons, but also non-linearly on the temperature of the impinging electrons \cite{blocker2016europa}. We therefore investigate the sensitivity of our results to the assumed value of $f_\mathrm{imp}$. We conduct two simulations in which the ionization rate of both $\mathrm{O_2}^+$ and $\mathrm{H_2O}^+$ is multiplied by 0.5 in the first one ($f_\mathrm{imp} = 10^{-6}~\mathrm{s}^{-1}$), and by 2 in the second one ($f_\mathrm{imp} = 4 \times 10^{-6}~\mathrm{s}^{-1}$). Both values are within (or close to) the range provided by \citeA{smyth2006europa}. As before, the assumed atmospheric model is number 3 from Section \ref{sec:oxygen emission ratio}, namely $\mathrm{O_2}$ and $\mathrm{H_2O}$ column densities $50\%$ of the values derived from \citeA{roth2021stable}, and an $\mathrm{H_2O}$ distribution with $\beta = 10$ and centered at the subsolar point. 

\begin{figure}
\centering
\noindent\includegraphics[width=0.5\textwidth]{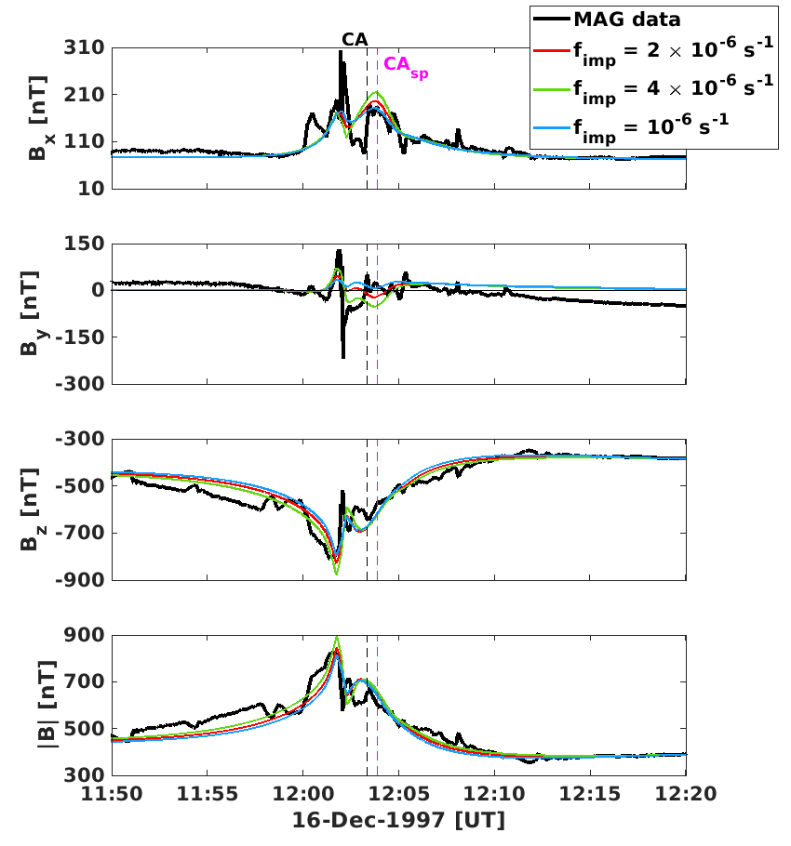}
\caption{Sensitivity analysis of the electron population. The black line indicates Galileo MAG data for the E12 flyby trajectory. Color coded are different simulated magnetic fields for various values the electron impact ionization rate $f_\mathrm{imp}$.}
\label{fig:zeus-R21-50-ionisation}
\end{figure}

Figure \ref{fig:zeus-R21-50-ionisation} presents the simulated magnetic field for these three scenarios. All the components of the magnetic field are perturbed at the location of the $\mathrm{H_2O}$ atmosphere, albeit at different amplitudes. The case with ionization rate $f_\mathrm{imp} = 4 \times 10^{-6}~\mathrm{s}^{-1}$ overestimates the local maximum due to the $\mathrm{H_2O}$ atmosphere in the $x$ component by $\sim$31 nT, whereas with $f_\mathrm{imp} = 10^{-6}~\mathrm{s}^{-1}$ the predicted $B_x$ only differs by $\sim$5 nT from the observations. The perturbations in $B_y$, both due to the plume and the $\mathrm{H_2O}$ around the subsolar point, are of larger amplitude for the case with $f_\mathrm{imp} = 4 \times 10^{-6}~\mathrm{s}^{-1}$, diverging the most from the observed field, especially after the plume occurrence. A similar pattern is evident in the $B_z$ component, where the model with $f_\mathrm{imp} = 4 \times 10^{-6}~\mathrm{s}^{-1}$ overestimates the  minimum due to the plume by $\sim$65 nT, whereas the other two cases only differ from the observed value by $\sim$16 nT. The local minimum in $B_z$ around closest approach is reproduced well by the three ionization rates. 

Our parameter study demonstrates that $\mathrm{O_2}$ and $\mathrm{H_2O}$ column densities reduced by $50\%$ relative to \citeA{roth2021stable} consistently match the amplitude and the location of the observed magnetic field perturbations. In other words, our simulations invariably require low column densities, but still within the error bars of the $r_\gamma$ profile of \citeA{roth2021stable}, to be in agreement with the MAG data. Most importantly, this conclusion holds after considering uncertainties in our atmospheric and MHD model, such as the location of the $\mathrm{H_2O}$ distribution and the electron impact ionization rate. 

Our results also show that variations of $\mathrm{O_2}$ and $\mathrm{H_2O}$ densities by a factor of 2 (Figure~\ref{fig:zeus-R21-75-50}) result in larger magnetic field perturbations than those due to an increase in the ionization rate by the same factor (Figure~\ref{fig:zeus-R21-50-ionisation}). This pattern suggests that, for our specific simulation of Galileo E12 flyby, the effect of electron impact ionization is weak, and thus, ion-neutral collisions play a dominant role in the overall plasma interaction.

\section{Summary and Conclusions}

In this work, we present new constraints on the density and spatial distribution of $\mathrm{O_2}$ and $\mathrm{H_2O}$ at Europa's atmosphere using a joint set of observations: HST spectral images of the trailing side of the moon, and Galileo magnetometer data for E12 flyby. We study the effect of a stable $\mathrm{H_2O}$ component concentrated around the subsolar point on the moon's plasma interaction. In addition, we perform a parameter study of the $\mathrm{H_2O}$ distribution and the electron impact ionization rate. 

We describe Europa's atmosphere with three neutral species: $\mathrm{O_2}$, $\mathrm{O}$, and $\mathrm{H_2O}$; and we obtain several distributions by progressively reducing the original $\mathrm{O_2}$ and $\mathrm{H_2O}$ column densities from \citeA{roth2021stable}. We take into account the uncertainty in the abundance of $\mathrm{O}$, ranging from its absence ($N_\mathrm{O} = 0$) to the upper limit provided by \citeA{roth2021stable} ($N_\mathrm{O} = 6 \times 10^{16}~\mathrm{m}^{-2}$), and we also consider the spatial variability of the stable $\mathrm{H_2O}$, as represented by the exponent $\beta$ of the cosine of the angle to the subsolar point. We find that several of the assumed abundances fit the observed $r_\gamma$ from HST within its error bars. Therefore, the emission ratio profile from \citeA{roth2021stable} only places conditional constraints on the column densities of $\mathrm{O_2}$ and $\mathrm{H_2O}$. Our simulated profiles also show that, as the degree of confinement of the stable $\mathrm{H_2O}$ around the subsolar point increases, less $\mathrm{H_2O}$ density is required in order to yield an oxygen emission ratio consistent with the observed profile.

In addition, we use a single-fluid MHD model to simulate the plasma interaction with Europa's atmosphere. Our results demonstrate that $N_\mathrm{O_2}$ $50\%$ and $ N_\mathrm{H_2O}$ between 50\% and 75\% from the values in \citeA{roth2021stable}, while strongly confining the stable $\mathrm{H_2O}$ with $\beta = 10$, jointly provide the best fit to both HST and MAG data. These percentages correspond to $N_\mathrm{O_2} = 1.24\times10^{18}~\mathrm{m}^{-2}$ and $N_\mathrm{H_2O}$ ranging from $1.47\times10^{19}~\mathrm{m}^{-2}$ to $2.22\times10^{19}~\mathrm{m}^{-2}$.

We show that the magnetic field fluctuations observed by Galileo after closest approach, mainly evident as a local maximum in $B_x$, are a signature of a confined $\mathrm{H_2O}$ atmosphere around the subsolar point. Furthermore, the parameter study demonstrates that our decreased densities perform well with a variety of $\mathrm{H_2O}$ and electron properties. As a consequence, a good agreement between MAG observations and the MHD simulations always requires low $\mathrm{O_2}$ and $\mathrm{H_2O}$ column densities, within the error bars of \citeA{roth2021stable}. 

Our findings are significant in a number of ways. We provide the first evidence of a localized persistent $\mathrm{H_2O}$ atmosphere concentrated around the subsolar point in Galileo magnetometer data, and we jointly limit its density by employing two independent datasets. Our derived constraints on the location and abundance of the $\mathrm{H_2O}$ distribution will help to understand the origin of such stable component. Finally, both JUICE \cite{grasset2013jupiter} and Europa Clipper missions \cite{howell2020nasa} will conduct several low-altitude passes above Europa's surface. Our results provide the mission teams with valuable information on the location and density of a stable $\mathrm{H_2O}$ atmosphere on the moon's trailing hemisphere. In-situ plasma and magnetic field measurements, particularly those in the subsolar region, will place additional observational constraints and refine our characterization of Europa's neutral atmosphere. 

\section{Open Research}

The ZEUS-MP code is publicly available and can be downloaded from \url{http://
www.netpurgatory.com/zeusmp.html}. The Galileo Magnetometer data were retrieved from the NASA Planetary Data System at GO-J-MAG-3-RDR-HIGHRES-V1.0 (doi: 10.17189/1519667). The location of the subsolar point was determined using the \texttt{solar\_system\_v0039.tm} meta-kernel provided by JPL.


\acknowledgments
This project that has received funding from the European Research Council (ERC) under the European Union’s Horizon 2020 research and innovation programme (grant agreement no. 884711). The numerical calculations have been performed on the CHEOPS Cluster of the University of Cologne, Germany. We thank Clarissa Willmes and Stefan Duling for their helpful advice on the usage of ZEUS-MP.


%
%



\bibliography{references}

%
%
%
%
%

\end{document}